\documentclass[aps,prb,10pt,english]{revtex4-1}
\usepackage[utf8]{inputenc}
\usepackage[letterpaper]{geometry}
\usepackage{xcolor}
\usepackage{babel}
\usepackage{amsmath}
\usepackage{amssymb}
\usepackage{graphicx}
\usepackage{setspace}
\PassOptionsToPackage{normalem}{ulem}
\usepackage{ulem}
\usepackage[unicode=true,
 bookmarks=false,
 breaklinks=false,pdfborder={0 0 1},backref=section,colorlinks=true]
 {hyperref}
\hypersetup{
 final,linkcolor=blue,citecolor=blue,filecolor=magenta,urlcolor=blue}

\DeclareTextSymbolDefault{\textquotedbl}{T1}

\begin{document}
\title{Exactly Thermalised Quantum Dynamics of the Spin-Boson Model coupled
to a Dissipative Environment}
\author{M. A. Lane$^{1}$, D. Matos$^{1}$, I. J. Ford$^{2}$, L. Kantorovich$^{1}$}
\affiliation{\singlespacing{}$^{1}$Department of Physics, King's College London, Strand, London,
WC2R 2LS, United Kingdom}
\affiliation{\singlespacing{}$^{2}$Department of Physics and Astronomy, University College London,
Gower Street, London, WC1E 6BT, United Kingdom}

\begin{abstract}
We present an application of the Extended Stochastic Liouville-von
Neumann equations (ESLN) method introduced earlier {[}PRB \textbf{95},
125124 (2017); PRB\textbf{ 97}, 224310 (2018){]} which describes the
dynamics of an exactly thermalised open quantum system reduced density
matrix coupled to a non-Markovian harmonic environment. Critically,
the combined system of the open system fully coupled to its environment
is thermalised at finite temperature using an imaginary time evolution
procedure before the application of real time evolution. This initialises
the combined system in the correct canonical equilibrium state rather
than being initially decoupled. Here we apply our theory to the spin-boson
Hamiltonian and develop a number of competing ESLN variants designed
to reduce the numerical divergence of the trace of the open system
density matrix. We find that a careful choice of the driving noises
is essential for improving numerical stability. We have also investigated
the effect of applying higher order numerical schemes for solving
stochastic differential equations, such as the Stratonovich-Heun scheme,
and concluded that stochastic sampling dominates convergence with
the improvement associated with the numerical scheme being less important
for short times but required for late times. To verify the method
and its numerical implementation, we first consider evolution under
a fixed Hamiltonian and show that the system either remains in, or
approaches, the correct canonical equilibrium state at long times.
Additionally, evolution of the open system under non-equilibrium Landau-Zener
(LZ) driving is considered and the asymptotic convergence to the LZ
limit was observed for vanishing system-environment coupling and temperature.
When coupling and temperature are non-zero, initially thermalising
the combined system at a finite time in the past was found to be a
better approximation of the true LZ initial state than starting in
a pure state.
\end{abstract}
\maketitle

\section{Introduction}

In open quantum systems, interactions between the system of interest
and its environment drive behaviours such as dissipation and decoherence
which are not found in isolation. This plays a strong role in quantum
computing \citep{shor1995scheme} and quantum thermodynamics \citep{weiss2012quantum}
where the ability of an open system to stay in a superposition of
states is desirable. However, the treatment of such systems is challenging
both analytically and numerically. Existing methods are typically
characterised by use of the reduced density matrix, obtained by taking
the partial trace over the environment variables of the full density
matrix. This began with the development of the Feynman-Vernon influence
functional formalism where the response of a linear bath is expressed
as a path integral over an infinite number of displaced harmonic oscillators
\citep{feynman2000theory}. Several techniques have since been developed,
including hierarchical equations of motion \citep{yan2004hierarchical,suess2014hierarchy},
stochastic Liouville-von Neumann equations (SLNs) \citep{diosi1998non,stockburger2001non,stockburger2002exact,stockburger2004simulating,mccaul2017partition},
stochastic Schrödinger equations \citep{orth2013nonperturbative}
and quasiadiabatic path integrals \citep{makri1995tensor}. Importantly,
none of these methods make the Markov assumption, where the environment
correlation times are taken to be negligibly short compared to the
characteristic timescales of the system of interest. This assumption
has the physical interpretation that any information dissipated from
the system to the environment will never be returned, i.e. the system-environment
coupling is memoryless. However, these methods do assume that the
system of interest and its environment are initially partitioned from
each other, that is, they are initially decoupled and thermalised
independently rather than as one combined system. This is fundamentally
unphysical, especially for driven systems where a partitioned state
is certainly not a good approximation of the correct initial thermal
state and leads to incorrect transient dynamics with the possibility
of the wrong asymptotic behaviour.

This is not the case for the recently proposed Extended Stochastic
Liouville-von Neumann equations (ESLN) method \citep{mccaul2017partition},
which builds on the earlier work of Graber, Schramm and Ingold \citep{grabert1988quantum}
and allows one to derive the equations of motion for the reduced density
matrix of an open quantum system without assuming a partitioned initial
state. It provides an exact, non-perturbative set of two stochastic
differential equations (SDEs): one in imaginary time that thermalises
the coupled system and the environment as a whole, and a second being
the typical stochastic Liouville-von Neumann equation (SLN) for the
open system. The thermalised state obtained at the end of an imaginary
time evolution becomes the initial state for the SLN so that sampling
over all manifestations of the noises leads to the exact dynamics
of the reduced density matrix of the open system starting in its thermal
state. Equations for observables can then be obtained in the usual
way. Crucially, the real time SLN dynamics is affected by the coupling
of the system to the environment during thermal preparation through
the correlation of the real time and imaginary time noises. This has
the natural interpretation that the preparation of the system may
influence any early time transient dynamics and perhaps even its asymptotic
behaviour at long times.

To simulate these stochastic differential equations, particular care
should be taken with the choice of numerical scheme and the manner
by which the coloured noises are generated. The latter point is not
trivial as the correlation functions in real time, imaginary time
and a cross-time correlation between them, must be satisfied with
sensible choices being made \citep{noise-generation-Matt-Dan-2020}.
It turns out that some allowed choices result in numerical instability
during the early time dynamics, even though the correlation functions
are fully satisfied. In our previous work \citep{mccaul2018driving},
a method for noise generation was proposed which we shall review and
further develop here, introducing a modified noise generation scheme
that diminishes the exponential growth of the trace of the density
matrix that seems to characterise these methods. This is the latest
in a series of proposals aimed at tackling this problem \citep{imai2015fmo,stockburger2019variance}.

To test the accuracy of the ESLN method, the spin-boson model will
be considered as the test-bed. It is typically the initial starting
model for any approach that deals with open quantum systems, due to
its relative simplicity while still exhibiting dissipative behaviour.
The model consists of a two level spin system surrounded by bosonic
degrees of freedom that describe the environment, and can naturally
be applied to qubits coupled to an environment \citep{duan1998reducing,costi2003entanglement,van2003engineering,kopp2007universal,cui2009non},
electronic energy transfer in biological systems \citep{imai2015fmo},
Josephson junctions \citep{makhlin2001quantum,liu2002theory,valenti2014switching},
cold atoms \citep{orth2010dynamics,orth2008dissipative} and solid-state
artificial atoms \citep{berns2008amplitude}. The spin-boson model
has already been considered previously by us in the context of ESLN
\citep{mccaul2018driving}; however, due to a recently discovered
implementation error, the numerical results were inaccurate. Here
we present further implementation development and update our numerical
results.

So, the purpose of the present paper is fourfold: (i) review and extend
existing methods of solving the dynamics of open quantum systems when
the density matrix is initialised in the correct canonical equilibrium
state; (ii) pay special attention to the generation of coloured Gaussian
noises for both real and imaginary time evolutions; (iii) examine
the convergence properties of two numerical schemes, one of which
uses Stratonovich calculus; and (iv) test the numerical behaviour
of different trace preserving forms of the ESLN and explain their
divergent behaviour in detail. In Section \ref{sec:theory} we briefly
review the ESLN \citep{mccaul2017partition} before moving on to the
spin-boson model. Section \ref{sec:noise_generation} presents the
schemes for noise generation along with techniques for reducing the
exponential growth of the trace, while Section \ref{sec:diff_forms_ESLE}
discusses various forms of ESLN including two trace preserving forms
obtained via a Girsanov transformation \citep{girsanov1960transforming,stockburger2004simulating}.
In Section \ref{sec:Stochastic-Differential-Equation} we discuss
schemes for solving the ESLN numerically using methods rooted in stochastic
calculus. Results of numerical simulations are given in Section \ref{sec:Results}
and the discussion and conclusions are presented in Section \ref{sec:Discussion-and-Conclusions}.

\section{Theory\label{sec:theory}}

\subsection{Extended Stochastic Liouville-von Neumann equations}

Following the influence functional formalism of Feynman and Vernon
\citep{feynman2000theory}, we consider the standard setup of an open
quantum system with coordinates $q$ and Hamiltonian $H_{q}$ that
may describe either an electronic or bosonic subsystem, or both, and
may depend explicitly on time. This system is coupled to its environment:
a heat bath of harmonic atoms $i$ with masses $m_{i}$, and a potential
energy that is quadratic in their displacement coordinates $\xi_{i}$.
The coupling between the open system and its environment is linear
in the environment coordinates but fully general in $q$, taking the
form $\xi_{i}f_{i}(q)$, with the set of $f_{i}(q)$ being arbitrary
functions of $q$. The full system Hamiltonian is thus 
\begin{equation}
H_{\textnormal{tot}}(q,\{\xi_{i}\},t)=H_{q}(q,t)+\sum_{i}\frac{p_{i}^{2}}{2m_{i}}+\frac{1}{2}\sum_{ij}\Lambda_{ij}\xi_{i}\xi_{j}-\sum_{i}\xi_{i}f_{i}(q),\label{eq:open_system_H}
\end{equation}
where $p_{i}$ are momenta coordinates canonical to $\xi_{i}$, and
$\Lambda_{ij}$ is the force constant matrix of the bath. A transformation
to normal modes then represents the bath as a set of non-interacting
harmonic oscillators. This is a more general form of the Caldeira-Leggett
Hamiltonian \citep{caldeira1983path} since the environment coupling
is a general function of $q$ rather than being strictly bilinear.

In typical studies, the open system and environment density matrix
is initialised in a partitioned state where the full system density
matrix $\boldsymbol{\rho}_{0}=\boldsymbol{\rho}_{\textnormal{tot}}(t_{0})$
is the tensor product of the open system density matrix $\boldsymbol{\rho}_{q}(t_{0})$
and that of its environment $\boldsymbol{\rho}_{\xi}(t_{0})$ at some
initial time $t_{0}$, 
\begin{equation}
\boldsymbol{\rho}_{0}=\boldsymbol{\rho}_{q}(t_{0})\otimes\boldsymbol{\rho}_{\xi}(t_{0}).\label{eq:full_initial}
\end{equation}
The more appropriate and useful initial state would be the one where
the open system and its environment are coupled and in thermal equilibrium.
This can be obtained via appropriate preparation of the canonical
equilibrium density matrix \citep{grabert1988quantum}, 
\begin{equation}
\boldsymbol{\rho}_{0}=\frac{1}{Z_{0}}e^{-\beta H_{0}},\label{eq:canonical_equilibrium}
\end{equation}
where $H_{0}=H_{\textnormal{tot}}(t_{0})$ is the initial Hamiltonian
of the combined open system and its environment, $Z_{0}=\text{Tr}\left(e^{-\beta H_{0}}\right)$
is the equilibrium partition function of the total system, and $\beta=1/k_{B}T$
is the inverse temperature.

Following the seminal work of Graber, Schramm and Ingold \citep{grabert1988quantum},
it was recently shown \citep{mccaul2017partition,mccaul2018driving}
that it is possible to thermalise the reduced density matrix of the
open system via a novel application of the influence functional formalism
in which the environment variables are integrated out for arbitrary
real time $t$. The resulting pair of SDEs describing the thermalisation
in imaginary time and subsequent dynamics in real time of the stochastic
reduced density matrix are known as the Extended Stochastic Liouville-von
Neumann equations (ESLN), with the evolution of the reduced density
matrix being driven by complex correlated Gaussian noises in both
cases. Expressing the equation of motion of the physical reduced density
matrix as an ensemble average over stochastic paths via a Hubbard-Stratonovich
transformation in this way is commonly referred to as stochastic unravelling
\citep{breuer2009stochastic,doi:10.1080/00268976.2018.1456685,moodley2009stochastic}.

Thermalisation is described by evolution in imaginary time $\tau$
of a density matrix $\overline{\boldsymbol{\rho}}(\tau)$ over the
domain $\tau\in[0,\beta\hbar]$ via 
\begin{equation}
-\hbar\frac{d\overline{\boldsymbol{\rho}}(\tau)}{d\tau}=\left(H_{q}(t_{0})+\sum_{i}\mu_{i}(\tau)f_{i}(q)\right)\overline{\boldsymbol{\rho}}(\tau),\label{eq:thermalisation}
\end{equation}
with $\overline{\boldsymbol{\rho}}(\tau)$ initialised in the unitary
state, $\overline{\boldsymbol{\rho}}(\tau=0)=\mathbb{{I}}$. The final
value of this evolution at $\tau=\beta\hbar$ corresponds to the equilibrium
density matrix, up to a normalisation constant which will be fixed
later. This is then used as the initial condition for the real-time
dynamics of the reduced density matrix which satisfies 
\begin{equation}
i\hbar\frac{d\boldsymbol{\rho}(t)}{dt}=\left[H_{q}(t),\boldsymbol{\rho}(t)\right]-\sum_{i}\left(\eta_{i}(t)\left[f_{i}(q),\boldsymbol{\rho}(t)\right]+\frac{\hbar}{2}\nu_{i}(t)\left\{ f_{i}(q),\boldsymbol{\rho}(t)\right\} \right)\label{eq:dynamics}
\end{equation}
where the square(curly) brackets represent the standard (anti-)commutators.

The functions $\eta_{i}(t)$, $\nu_{i}(t)$ and $\mu_{i}(\tau)$ are
the driving complex Gaussian noises, distributed via the multivariate
Gaussian 
\[
\mathcal{W}\left[\left\{ \mu_{i}\right\} ,\left\{ \eta_{i}\right\} ,\left\{ \nu_{i}\right\} \right]=\mathcal{N}\exp\left\{ -\frac{1}{2}\left[\int_{0}^{t}dt'\int_{0}^{t}dt''\boldsymbol{z}_{1}^{T}(t')\boldsymbol{\Sigma}^{11}(t'-t'')\boldsymbol{z}_{1}(t'')\right.\right.
\]
\begin{equation}
\left.\left.+2\int_{0}^{t}dt'\int_{0}^{\beta\hbar}d\tau\boldsymbol{z}_{1}^{T}(t')\boldsymbol{\Sigma}^{12}(t',\tau)\boldsymbol{z}_{2}(\tau)+\int_{0}^{\beta\hbar}d\tau\int_{0}^{\beta\hbar}d\tau'\boldsymbol{z}_{2}^{T}(\tau')\boldsymbol{\Sigma}^{22}(\tau-\tau')\boldsymbol{z}_{2}(\tau')\right]\right\} ,\label{eq:Gaussian}
\end{equation}
and arising from the application of a two-time Hubbard-Stratonovich
transformation \citep{hubbard1959calculation,stockburger2002exact,mccaul2017partition}
to the environment influence functional. Here $\mathcal{N}$ is a
normalisation constant, $\boldsymbol{z}_{1}=\left(\left\{ \eta_{i}\right\} \:\left\{ \eta_{i}^{*}\right\} \:\left\{ \nu_{i}\right\} \:\left\{ \nu_{i}^{*}\right\} \right)^{T}$
and $\boldsymbol{z}_{2}=\left(\left\{ \mu_{i}\right\} \:\left\{ \mu_{i}^{*}\right\} \right)^{T}$
are the vector noises, and the $\Sigma^{ij}$ are time dependent matrices
to be discussed shortly.

The physical reduced density matrix is obtained by the average $\langle\ldots\rangle$
of an ensemble of stochastic reduced density matrices, taken over
the noises with the multivariate Gaussian weighting given above. In
particular, the average at the end of imaginary time evolution yields
the exact thermalised initial state of the real time evolution, that
is, $\boldsymbol{\rho}^{ph}\left(t_{0}\right)=\mathcal{\mathbb{N}}\langle\boldsymbol{\overline{\rho}}(\beta\hbar)\rangle\equiv\mathcal{\mathbb{N}}\langle\boldsymbol{\rho}\left(t_{0}\right)\rangle$.
Here, $\mathbb{N}$ is a time-independent pre-factor that is to be
fixed \citep{mccaul2017partition,mccaul2018driving} after sampling
using the condition $\text{Tr\ensuremath{\left(\boldsymbol{\rho}^{ph}\left(t\right)\right)}}=\mathcal{\mathbb{N}}\text{Tr}\left(\langle\boldsymbol{\rho}(t)\rangle\right)=1$.
In practice this can be done at any time including $t_{0}$, so the
physical density matrix is obtained by taking $\mathbb{N}=1/\text{Tr}\left(\langle\boldsymbol{\rho}(t_{0})\rangle\right)$.

The blocks of the matrix $\boldsymbol{\Sigma}=\left(\begin{array}{cc}
\boldsymbol{\Sigma}^{11} & \boldsymbol{\Sigma}^{12}\\
\boldsymbol{\Sigma}^{21} & \boldsymbol{\Sigma}^{22}
\end{array}\right)$ in the Gaussian of Eq. \eqref{eq:Gaussian} are defined such that
the corresponding elements of its inverse are equal to the appropriate
noise correlation functions (given below). Only correlation functions
between noises $\eta_{i}(t)$, $\nu_{i}(t)$ and $\mu_{i}(\tau)$
are needed; other correlation functions involving complex conjugated
noises can be ignored \citep{mccaul2018driving}. It is important
to note that each realisation of these noises will produce a unique
trajectory describing an initial thermalised stochastic density matrix
and its subsequent real time dynamics, with the physical density matrix
obtained by stochastic averaging over a sufficiently large sample
of such realisations. This has the pleasingly intuitive interpretation
of averaging over all possible behaviours of the bath, reminiscent
of the direct link to the sum over all possible paths in the path
integral representation, only now this sum is replaced by the stochastic
average over environmental noises.

\subsection{Noise Correlation Functions}

The noises are defined by their site dependent correlation functions,
\begin{equation}
\langle\eta_{i}(t)\eta_{j}(t^{\prime})\rangle=\frac{\hbar}{\sqrt{m_{i}m_{j}}}\sum_{\lambda}\frac{e_{\lambda i}e_{\lambda j}}{2\omega_{\lambda}}\coth\left(\frac{1}{2}\beta\hbar\omega_{\lambda}\right)\cos\left(\omega_{\lambda}t\right),
\end{equation}
\begin{equation}
\langle\eta_{i}(t)\nu_{j}(t^{\prime})\rangle=-\frac{2i\Theta(t-t^{\prime})}{\sqrt{m_{i}m_{j}}}\sum_{\lambda}\frac{e_{\lambda i}e_{\lambda j}}{2\omega_{\lambda}}\coth\left(\frac{1}{2}\beta\hbar\omega_{\lambda}\right)\sin\left(\omega_{\lambda}t\right),
\end{equation}
\begin{equation}
\langle\eta_{i}(t)\mu_{j}(\tau)\rangle=-\frac{\hbar}{\sqrt{m_{i}m_{j}}}\sum_{\lambda}\frac{e_{\lambda i}e_{\lambda j}}{2\omega_{\lambda}}\frac{\cosh\left(\frac{1}{2}\beta\hbar\omega_{\lambda}-i\omega_{\lambda}\left(t-i\tau\right)\right)}{\sinh\left(\frac{1}{2}\beta\hbar\omega_{\lambda}\right)},
\end{equation}
\begin{equation}
\langle\mu_{i}(\tau)\mu_{j}(\tau^{\prime})\rangle=\frac{\hbar}{\sqrt{m_{i}m_{j}}}\sum_{\lambda}\frac{e_{\lambda i}e_{\lambda j}}{2\omega_{\lambda}}\left[\coth\left(\frac{1}{2}\beta\hbar\omega_{\lambda}\right)\cosh\left(\omega_{\lambda}\tau\right)-\sinh\left(\omega_{\lambda}\tau\right)\right]
\end{equation}
\begin{equation}
\langle\nu_{i}(t)\nu_{j}(t^{\prime})\rangle=\langle\nu_{i}(t)\mu_{j}(\tau)\rangle=0,\label{eq:non-needed}
\end{equation}
where $\Theta(t)$ is the Heaviside step function. Here, the $e_{\lambda}$
are the eigenvectors of the bath dynamical matrix, $D_{ij}=\Lambda_{ij}/\sqrt{m_{i}m_{j}}$,
with eigenvalues $\omega_{\lambda}^{2}$.

In the standard SLN without any thermalisation, there would be no
$\mu$ noise and no $\eta-\mu$ correlation. This is indicative of
the neglected information inherent in initializing the system in a
partitioned state. In the thermalised ESLN, thermalisation leads to
entanglement between the system of interest and its environment, manifested
in the $\eta-\mu$ cross correlation, which may persist after thermalisation
during the real time dynamics. At first glance this may seem strange,
since the cross-correlation between real and imaginary times refers
to two intrinsically \textit{different }time coordinates. Regardless,
the noises are auxiliary variables introduced by the application of
a two-time Hubbard-Stratonovich transformation; they do not have physical
meaning by themselves. Similarly, components of the stochastic density
matrix are simply mathematical degrees of freedom from the perspective
of the correlations functions, describing a random trajectory first
along the imaginary coordinate $\tau$ and second along the real coordinate
$t$, with the particular realisation of the latter depending on the
final realisation of the former. The physical density matrix is obtained
after averaging over these realisations, with each realisation being
a different stochastic quantum trajectory.

The general ESLN, Eqs. \eqref{eq:thermalisation} and \eqref{eq:dynamics},
requires three noises $\eta_{i}$, $\nu_{i}$ and $\mu_{i}$ per lattice
site $i$. In normal mode representation $\lambda$ the correlation
matrices are diagonalised. Next, by assuming that the system variable
dependence of the system-environment coupling, $-\sum_{\lambda}f_{\lambda}(q)\xi_{\lambda}$,
is the same for each mode up to a scaling factor, $f_{\lambda}(q)=c_{\lambda}f(q)$,
the set of noise terms can be reduced from three per site down to
only three \citep{mccaul2018driving}. For example, taking the $\eta_{i}\rightarrow\eta_{\lambda}$
noise, the $\eta_{i}$ term in Eq. \eqref{eq:dynamics} becomes 
\begin{equation}
\sum_{i}\eta_{i}(t)\left[f_{i}(q),\boldsymbol{\rho}(t)\right]\ \rightarrow\ \eta(t)\left[f(q),\boldsymbol{\rho}(t)\right]
\end{equation}
with $\eta(t)=\sum_{\lambda}c_{\lambda}\eta_{\lambda}(t)$ being a
new Gaussian noise. The $\eta-\eta$ correlation function is then
\begin{equation}
\langle\eta(t)\eta(t^{\prime})\rangle=\hbar\sum_{\lambda}\frac{c_{\lambda}^{2}}{2\omega_{\lambda}}\coth\left(\frac{1}{2}\beta\hbar\omega_{\lambda}\right)\cos\left(\omega_{\lambda}\left(t-t^{\prime}\right)\right),
\end{equation}
where the sum over environmental modes can be replaced by an integration
over frequency in the continuum limit, 
\[
\sum_{\lambda}\frac{c_{\lambda}^{2}}{2\omega_{\lambda}}\ldots\ \rightarrow\ \int_{0}^{\infty}\frac{d\omega}{\pi}\left[\pi\sum_{\lambda}\frac{c_{\lambda}^{2}}{2\omega_{\lambda}}\delta\left(\omega-\omega_{\lambda}\right)\right]\ldots=\int_{0}^{\infty}\frac{d\omega}{\pi}J(\omega)\ldots.
\]
Here, $J(\omega)$ is the spectral density of the environment, taken
in this work to be the Drude spectral density, 
\begin{equation}
J(\omega)=\alpha\omega\left[1+\left(\frac{\omega}{\omega_{c}}\right)^{2}\right]^{-2},
\end{equation}
where $\alpha$ is proportional to the squares of the $c_{\lambda}$
coefficients and so parameterises the effective coupling strength
between the system and environment. $\omega_{c}$ is the Drude-Lorentz
cut-off frequency which ensures that the density goes smoothly to
zero as $\omega$ becomes large \citep{breuer2002theory,cui2009non}.

Just as for the $\eta$ noise, the sets of $\nu_{i}$ and $\mu_{i}$
noises may be reduced to only a single $\nu$ and $\mu$ Gaussian
noise, and the sums over $i$ in Eqs. \eqref{eq:thermalisation} and
\eqref{eq:dynamics} are completely removed. The correlation functions
for these three reduced noises are 
\begin{equation}
\langle\eta(t)\eta(t^{\prime})\rangle=\hbar\int_{0}^{\infty}\frac{d\omega}{\pi}J(\omega)\coth\left(\frac{1}{2}\beta\hbar\omega\right)\cos\left(\omega\left(t-t^{\prime}\right)\right)\equiv K_{\eta\eta}(t-t'),\label{eq:kernel_etaeta}
\end{equation}
\begin{equation}
\langle\eta(t)\nu(t^{\prime})\rangle=-2i\hbar\Theta(t-t')\int_{0}^{\infty}\frac{d\omega}{\pi}J(\omega)\sin\left(\omega\left(t-t^{\prime}\right)\right)\equiv K_{\eta\nu}(t-t'),\label{eq:kernel_etanu}
\end{equation}
\begin{equation}
\langle\eta(t)\mu(\tau)\rangle=-\hbar\int_{0}^{\infty}\frac{d\omega}{\pi}J(\omega)\frac{\cosh\left(\frac{1}{2}\beta\hbar\omega-i\omega\left(t-i\tau\right)\right)}{\sinh\left(\frac{1}{2}\beta\hbar\omega\right)}\equiv K_{\eta\mu}(t,\tau),\label{eq:kernel_etamu}
\end{equation}
\begin{equation}
\langle\mu(\tau)\mu(\tau^{\prime})\rangle=\hbar\int_{0}^{\infty}\frac{d\omega}{\pi}J(\omega)\left[\cosh\left(\omega\left(\tau-\tau'\right)\right)\coth\left(\frac{1}{2}\beta\hbar\omega\right)-\sinh\left(\omega\left(\tau-\tau^{\prime}\right)\right)\right]\equiv K_{\mu\mu}(\tau-\tau'),\label{eq:kernel_mumu}
\end{equation}
\begin{equation}
\langle\nu(t)\nu(t^{\prime})\rangle=\langle\nu(t)\mu(\tau)\rangle=0,\quad\forall t,t^{\prime},\tau,\label{eq:zero_correlations}
\end{equation}
where we have defined so-called physical kernels on the right hand
sides. Note that Eq. \eqref{eq:zero_correlations} is possible because
the noises are complex valued. Correspondingly, Eqs. \eqref{eq:thermalisation}
and \eqref{eq:dynamics} are simplified as 
\begin{equation}
-\hbar\frac{d\overline{\boldsymbol{\rho}}(\tau)}{d\tau}=\left(H_{q}(t_{0})+\mu(\tau)f(q)\right)\overline{\boldsymbol{\rho}}(\tau),\label{eq:thermalisation-1}
\end{equation}
\begin{equation}
i\hbar\frac{d\boldsymbol{\rho}(t)}{dt}=\left[H_{q}(t),\boldsymbol{\rho}(t)\right]-\eta(t)\left[f(q),\boldsymbol{\rho}(t)\right]-\frac{\hbar}{2}\nu(t)\left\{ f(q),\boldsymbol{\rho}(t)\right\} .\label{eq:dynamics-1}
\end{equation}
Note that formally Eq. \eqref{eq:dynamics-1} coincides with the SLN
dynamics. The important difference here lies in the cross-correlation
with the imaginary time dynamics associated with thermalisation, and
the use of the final result of each $\overline{\boldsymbol{\rho}}(\tau)$
as the initial condition for each $\boldsymbol{\rho}(t)$.

\subsection{Spin-Boson Model}

Thus far, the system Hamiltonian $H_{q}$ has been kept fully general,
as has the form of the system-environment coupling, $f(q)$. The spin-boson
Hamiltonian for a generic two-state system,
\begin{equation}
H_{q}(t)=\frac{1}{2}\hbar\Delta(t)\sigma_{x}+\frac{1}{2}\hbar\epsilon(t)\sigma_{z}=\frac{1}{2}\hbar\Delta(t)\left(\lvert0\rangle\langle1\rvert+\lvert1\rangle\langle0\rvert\right)+\frac{1}{2}\hbar\epsilon(t)\left(\lvert0\rangle\langle0\rvert-\lvert1\rangle\langle1\rvert\right),\label{eq:spin_boson_H}
\end{equation}
is a good model in which to confirm the efficacy of the ESLN. Here
$\sigma_{x}$ and $\sigma_{z}$ are the standard Pauli spin matrices
with $\sigma_{x}$ flipping the spin from one state to the other with
tunnelling strength $\Delta(t)$, and $\sigma_{z}$ biasing the states
with magnitude $\epsilon(t)$. The system-bath coupling is just $\sigma_{z}$
so that Eqs. \eqref{eq:thermalisation-1} and \eqref{eq:dynamics-1}
become 
\begin{equation}
-\hbar\frac{d\overline{\boldsymbol{\rho}}(\tau)}{d\tau}=\left(H(t_{0})+\mu(\tau)\sigma_{z}\right)\overline{\boldsymbol{\rho}}(\tau),\label{eq:ESLE_thermalisation}
\end{equation}
\begin{equation}
i\hbar\frac{d\boldsymbol{\rho}(t)}{dt}=\left[H(t),\boldsymbol{\rho}(t)\right]-\eta(t)\left[\sigma_{z},\boldsymbol{\rho}(t)\right]-\frac{\hbar}{2}\nu(t)\left\{ \sigma_{z},\boldsymbol{\rho}(t)\right\} .\label{eq:ESLE-method-1}
\end{equation}
The total system is first jointly thermalised using Eq. \eqref{eq:ESLE_thermalisation}
so that at $\tau=\beta\hbar$ the sample average produces the equilibrium
state Eq. \eqref{eq:canonical_equilibrium}. Each stochastic $\boldsymbol{\rho}(t)$
is then initialised at $t_{0}$ in the corresponding equilibrium state
$\overline{\boldsymbol{\rho}}(\beta\hbar)$ and evolved in real time
according to Eq. \eqref{eq:ESLE-method-1}. Finally, the normalisation
factor $\mathbb{N}$ is determined and the full physical reduced density
matrix becomes completely defined.

In this work two simple tests for the dynamics are discussed. First,
we consider equilibrium evolution with constant driving whereby the
system decays towards the thermal state if initialised elsewhere or
remains unperturbed if initialised in the thermal state. And second,
a linear driving after some initial time $t_{0}$ of the form $\epsilon(t)=\kappa t$
with constant $\Delta$ is investigated, known as the Landau-Zener
sweep \citep{zener1932non}. Importantly, for an isolated spin being
linearly driven from $\epsilon(-\infty)=-\infty$ to $\epsilon(+\infty)=\infty$
at zero temperature starting in the ground state $\lvert1\rangle$,
or $\rho_{ij}\left(-\infty\right)=\delta_{i1}\delta_{j1}$, the survival
probability as $t\rightarrow\infty$ is \citep{zener1932non,wittig2005landau,rojo2010matrix,saito2007dissipative,orth2013nonperturbative,nalbach2009landau}

\begin{equation}
P_{LZ}=\exp\left\{ \frac{\pi\Delta^{2}}{2\hbar\kappa}\right\} \label{eq:survival}
\end{equation}
which corresponds to an asymptotic mean $z-$spin of 
\begin{equation}
\langle\sigma_{z}\rangle_{LZ}=2\exp\left\{ -\frac{\pi\Delta^{2}}{2\hbar\kappa}\right\} -1.\label{eq:LZlimit}
\end{equation}

Though this result was originally derived for an isolated spin, it
has since been shown that the same asymptotic behavior is valid for
a dissipative spin coupled to a harmonic environment at zero temperature,
where coupling is provided entirely via $\sigma_{z}$ \citep{wubs2006gauging,saito2007dissipative,rojo2010matrix,orth2013nonperturbative}.
This correspondence breaks down if the initial condition is not the
ground state $\lvert1\rangle$ in the infinite past, or for non-zero
temperature.

Finally, using Eqs. \eqref{eq:ESLE_thermalisation} and \eqref{eq:ESLE-method-1}
for the spin-boson Hamiltonian it is straightforward to derive coupled
SDEs for the $x$, $y$ and $z$-spins and also for the trace, $\text{Tr}\left({\boldsymbol{\rho}(t)}\right)$,
\begin{equation}
\hbar\frac{d\sigma_{x}(t)}{dt}=-\left[\epsilon(t)-2\eta(t)\right]\sigma_{y}(t)\label{eq:xspin}
\end{equation}
\begin{equation}
\hbar\frac{d\sigma_{y}(t)}{dt}=-\Delta\sigma_{z}(t)+\left[\epsilon(t)-2\eta(t)\right]\sigma_{x}(t)\label{eq:yspin}
\end{equation}
\begin{equation}
\hbar\frac{d\sigma_{z}(t)}{dt}=\Delta\sigma_{y}(t)+i\nu(t)\,\text{Tr}\left({\boldsymbol{\rho}(t)}\right)\label{eq:zspin}
\end{equation}
\begin{equation}
\hbar\frac{d\,\text{Tr}\left({\boldsymbol{\rho}}(t)\right)}{dt}=i\nu(t)\sigma_{z}(t),\label{eq:trace_evolution1}
\end{equation}
where the last equation is obtained by taking the trace of Eq. \eqref{eq:ESLE-method-1}.
To be clear, here $\sigma_{i}$ without time is just the usual Pauli
spin matrix, while $\sigma_{i}(t)=\text{Tr}{\left(\sigma_{i}\boldsymbol{\rho}(t)\right)}$
is the quantum average using a single realisation of the density matrix,
and 
\begin{equation}
\langle\sigma_{i}(t)\rangle=\text{Tr}{\left(\sigma_{i}\boldsymbol{\rho}^{ph}(t)\right)}=\text{Tr}{\left(\sigma_{i}\frac{\langle\boldsymbol{\rho}(t)\rangle}{\text{Tr}{\left(\langle\boldsymbol{\rho}(t_{0})\rangle\right)}}\right)}\label{eq:average-sigma}
\end{equation}
 is the quantum average using the physical density matrix obtained
after stochastic averaging and normalisation.

\section{Noises \label{sec:noise_generation}}

\subsection{Noise Generation Scheme}

Compared to the SLN, the noises in the ESLN have the additional complexity
of an extra coloured noise $\mu$ with its own time coordinate $\tau$,
introducing cross-time correlations \citep{mccaul2017partition}.
Adopting the notation for the noises used in \citep{mccaul2018driving},
the correlation functions for the spin-boson Hamiltonian reduce to
Eqs. \eqref{eq:kernel_etaeta}-\eqref{eq:zero_correlations}. These
correlation functions act as constraints on any noise generated, but
the noises are not uniquely defined by them. This provides some freedom
in specifying the generation procedure, as long as the correlation
functions are satisfied.

Decomposing each noise into its orthogonal components such that each
component is correlated with only one other component, and denoting
the correlations between components with subscripts, the noises can
be written as 
\begin{equation}
\eta(t)=\eta_{\eta}(t)+\eta_{\nu}(t)+\eta_{\mu}(t)\label{eq:eta_noise}
\end{equation}
\begin{equation}
\nu(t)=\nu_{\eta}(t)\label{eq:nu_noise}
\end{equation}
\begin{equation}
\mu(\tau)=\mu_{\mu}(\tau)+\mu_{\eta}(\tau).\label{eq:mu_noise}
\end{equation}
Explicitly, this means that $\eta_{\nu}$ is only correlated with
$\nu_{\eta}$, with equivalent products for other orthogonal pairs.
This orthogonality can be achieved by expressing each component as
a convolution of an unknown time-function $G$ (to be called a filtering
kernel) with a sum of real valued white noises, satisfying
\begin{equation}
\langle x_{i}(t)x_{j}(t^{\prime})\rangle=\delta_{ij}\delta(t-t^{\prime})
\end{equation}
\begin{equation}
\langle\overline{x}_{i}(\tau)\overline{x}_{j}(\tau^{\prime})\rangle=\delta_{ij}\delta(\tau-\tau^{\prime})
\end{equation}
\begin{equation}
\langle x_{i}(t)\overline{x}_{j}(\tau)\rangle=0\quad\text{for}\quad\forall i,j.
\end{equation}
Here $x_{i}(t)$ and $\overline{x}_{i}(\tau)$ refer to a white noise
in real and imaginary time, respectively. The convolutions thus take
the form 
\begin{equation}
\eta_{\eta}(t)=\int_{-\infty}^{\infty}dt^{\prime}G_{\eta\eta}(t-t^{\prime})x_{1}(t^{\prime})\label{eq:noise-eta-eta}
\end{equation}
\begin{equation}
\eta_{\nu}(t)=\int_{-\infty}^{\infty}dt^{\prime}G_{\eta\nu}(t-t^{\prime})\left[x_{2}(t^{\prime})+ix_{3}(t^{\prime})\right]\label{eq:noise-eta-nu}
\end{equation}
\begin{equation}
\eta_{\mu}(t)=\int_{0}^{\beta\hbar}d\tau G_{\eta\mu}(t,\tau)\left[\overline{x}_{2}(\tau)+i\overline{x}_{3}(\tau)\right]\label{eq:noise-eta-mu}
\end{equation}
\begin{equation}
\nu_{\eta}(t)=\int_{-\infty}^{\infty}dt^{\prime}G_{\nu\eta}(t-t^{\prime})\left[x_{3}(t^{\prime})+ix_{2}(t^{\prime})\right]\label{eq:noise-nu-eta}
\end{equation}
\begin{equation}
\mu_{\mu}(\tau)=\int_{-\beta\hbar}^{\beta\hbar}d\tau^{\prime}G_{\mu\mu}(\tau-\tau^{\prime})\overline{x}_{1}(\tau^{\prime})\label{eq:noise-mu-mu}
\end{equation}
\begin{equation}
\mu_{\eta}(\tau)=\int_{0}^{\beta\hbar}d\tau'G_{\mu\eta}(\tau-\tau')[\overline{x}_{3}(\tau')+i\overline{x}_{2}(\tau')],\label{eq:noise-mu-eta}
\end{equation}
from which it is straightforward to show that the expectation values
of component pairs correspond to the appropriate correlation functions,
e.g. $\langle\eta(t)\nu(t^{\prime})\rangle=\langle\eta_{\nu}(t)\nu_{\eta}(t^{\prime})\rangle$.
The choice of each $G$ is made by equating the expectation values
of the noises to the appropriate physical kernels, $K\left(t\right)$,
Eqs. \eqref{eq:kernel_etaeta}-\eqref{eq:zero_correlations}, and
taking Fourier transforms (indicated by the tilde) where appropriate
to obtain 
\begin{equation}
\tilde{G}_{\eta\eta}(\omega)=\sqrt{\tilde{K}_{\eta\eta}(\omega)}\label{eq:filter-eta-eta}
\end{equation}
\begin{equation}
\tilde{G}_{\eta\nu}(\omega)=\tilde{G}_{\nu\eta}\left(-\omega\right)=\sqrt{-\frac{i}{2}\tilde{K}_{\eta\nu}(\omega)}\label{eq:filter-eta-nu}
\end{equation}
\begin{equation}
\tilde{G}_{\mu\mu}(\omega)=\sqrt{\tilde{K}_{\mu\mu}(\omega)}\label{eq:filter-mu-mu}
\end{equation}
\begin{equation}
G_{\eta\mu}(t,\tau)=-\frac{i}{2}K_{\eta\mu}(t-i\tau),\label{eq:filter-eta-mu}
\end{equation}
with the remaining filtering kernel given by a delta function $G_{\mu\eta}\left(\tau\right)=\delta\left(\tau\right)$.
Note that in our previous work \citep{mccaul2018driving} we used
$G_{\nu\eta}\left(t\right)=\delta\left(t\right)$ instead of Eq. \eqref{eq:filter-eta-nu},
which we have found leads to much less stable dynamics \citep{noise-generation-Matt-Dan-2020}.
The noises can then be obtained by applying the convolution theorem
to Eqs. \eqref{eq:noise-eta-eta}-\eqref{eq:noise-mu-mu} before taking
the inverse Fourier transform.

\subsection{Variance Reduction Technique\label{sec:variance_reduction}}

From the equation of motion for the trace, Eq. \eqref{eq:trace_evolution1},
and given that $\nu$ is complex valued, it is found that the trace
can grow exponentially in time \citep{stockburger2019variance}, requiring
punitively large sampling for convergence. Recent proposals to optimise
the noise generation method \citep{imai2015fmo,stockburger2019variance}
have managed to reduce this growth by many orders of magnitude, though
here we present a much simpler method of exploiting the relative magnitudes
of correlated pairs of orthogonal noises such that their correlation
functions do not change.

Since the noise components are orthogonal, the correlation functions
depend only on the two appropriate components, e.g, $K_{\eta\nu}(t-t^{\prime})=\left\langle \eta_{\nu}(t)\nu_{\eta}(t^{\prime})\right\rangle $,
so $\nu_{\eta}$ can be multiplied and $\eta_{\nu}$ divided by the
same factor without modifying the correlation, and equivalently for
$K_{\eta\mu}$.To accomplish this, we define the scaling factors 
\begin{equation}
a_{\mu\eta}=\sqrt{r_{\mu\eta}}\sqrt{\frac{\frac{1}{M}\sum_{m=0}^{M}\left|\mu_{\eta}(\tau_{m})\right|}{\max_{n}\left|\eta_{\mu}(t_{n})\right|}}\label{eq:scaling-eta-mu}
\end{equation}
\begin{equation}
b_{\nu\eta}=\sqrt{r_{\nu\eta}}\sqrt{\frac{\sum_{n=0}^{N}\left|\nu_{\eta}(t_{n})\right|}{\sum_{n=0}^{N}\left|\eta_{\nu}(t_{n})\right|}},\label{eq:scale_eta_nu}
\end{equation}
where $M=\beta\hbar/d\tau$ and $N=t_{max}/dt$ are the number of
real and imaginary time steps, respectively, with $\tau_{m}=md\tau$
and $t_{n}=ndt$, while $r_{\mu\eta}$ and $r_{\nu\eta}$ are the
desired average ratios of the relative components of the noises over
a single realisation. The desired new noises are thus obtained by simply
rescaling the components as $\eta_{\mu}^{new}=a_{\mu\eta}\eta_{\mu}$
and $\mu_{\eta}^{new}=\mu_{\eta}/a_{\mu\eta}$, and $\eta_{\nu}^{new}=b_{\nu\eta}\eta_{\nu}$
and $\nu_{\eta}^{new}=\nu_{\eta}/b_{\nu\eta}$. Here, the maximum
absolute value of $\eta_{\mu}$ rather than the average over its realisation
is used in Eq. \eqref{eq:scaling-eta-mu} since $\eta_{\mu}$ rapidly
attenuates with time. This ensures that the typical magnitude of features
in $\eta_{\mu}$ and $\mu_{\eta}$ are scaled, making it possible
to control the spread of initial values for the real time dynamics
by reducing the variance of thermalisation trajectories.

For example, for $r_{\nu\eta}=1$, the average magnitudes of $\eta_{\nu}^{new}$
and $\nu_{\eta}^{new}$ over a realisation are approximately equal.
Alternatively, $r_{\nu\eta}$ can be chosen to reduce the variance
of $\text{Tr}\left({\boldsymbol{\rho}(t)}\right)$ by reducing the
magnitude of $\nu$ close to zero. However, in Section \ref{subsec:noise_and_convergence}
we will show that taking this limit is not desirable as $\operatorname{Im}[\eta_{\nu}]$
grows with $r_{\nu\eta}$, resulting in numerical instability.

\section{Different forms of the ESLN\label{sec:diff_forms_ESLE}}

From Eq. \eqref{eq:trace_evolution1}, it is clear that the dynamics
of each stochastic $\boldsymbol{\rho}$ is not trace-preserving. This
can lead to exponential blow-up \citep{davila2005numerical,higham2001algorithmic}
of the trace and requires punitively large sample size for convergence.
One way of enforcing trace preservation is to instead consider the
trace-normalized density matrix, $\tilde{\boldsymbol{\rho}}(t)=\boldsymbol{\rho}(t)/\text{Tr}\left(\boldsymbol{\rho}(t)\right)$,
satisfying \citep{stockburger2004simulating}
\begin{equation}
i\hbar\frac{d\tilde{\boldsymbol{\rho}}(t)}{dt}=\left[H(t),\tilde{\boldsymbol{\rho}}(t)\right]-\eta(t)\left[\sigma_{z},\tilde{\boldsymbol{\rho}}(t)\right]-\frac{\hbar}{2}\nu(t)\left\{ \sigma_{z}-\sigma(t),\tilde{\boldsymbol{\rho}}(t)\right\} ,\label{eq:trace_preserving}
\end{equation}
where we have introduced the guide spin 
\begin{equation}
\sigma(t)=\frac{\text{Tr}\left(\sigma_{z}\boldsymbol{\rho}(t)\right)}{\text{Tr}\left(\boldsymbol{\rho}(t)\right)}=\text{Tr}\left(\sigma_{z}\tilde{\boldsymbol{\rho}}(t)\right).\label{eq:guide_spin-1}
\end{equation}

Simulating this normalised $\tilde{\boldsymbol{\rho}}(t)$ still requires
knowledge of the original $\text{Tr}\left(\boldsymbol{\rho}(t)\right)$
to perform the required statistical averaging since $\boldsymbol{\rho}^{ph}(t)=\langle\boldsymbol{\rho}(t)\rangle=\langle\tilde{\boldsymbol{\rho}}(t)\text{Tr}\left(\boldsymbol{\rho}(t)\right)\rangle$.
It is possible to overcome this problem via a transformation that
enforces trace preservation for each realisation while preserving
the original ensemble mean \citep{ghirardi1990markov,gatarek1991continuous,diosi1998non,stockburger2001non,stockburger2002exact},
i.e. $\boldsymbol{\rho}^{{ph}}(t)=\langle\tilde{\boldsymbol{\rho}}(t)\rangle$.
Such a transformation of the probability measure, $\mathcal{W}\rightarrow\mathcal{W}^{\prime}$,
is called a Girsanov transformation, where both the transformed and
the original measures give rise to identical observables \citep{stockburger2004simulating,gardiner2009stochastic,kloeden2012numerical,lawler2018introduction}.
That is, 
\begin{equation}
\boldsymbol{\rho}^{ph}(t)=\langle\boldsymbol{\rho}(t)\rangle_{\mathcal{{W}}}=\langle\tilde{\boldsymbol{\rho}}(t)\rangle_{\mathcal{{W}^{\prime}}},\label{eq:rho_girsanov}
\end{equation}
where $\langle\ldots\rangle_{\mathcal{{W}}}=\int d\boldsymbol{z}_{1}d\boldsymbol{z}_{2}\mathcal{W}\left[\boldsymbol{z}_{1},\boldsymbol{z}_{2}\right]\ldots$
denotes the ensemble average over noises $\boldsymbol{z}_{1}=\left(\eta\:\eta^{*}\:\nu\:\nu^{*}\right)^{T}$
and $\boldsymbol{z}_{2}=\left(\mu\:\mu^{*}\right)^{T}$ drawn from
the original Gaussian distribution $\mathcal{W}\left[\boldsymbol{z}_{1},\boldsymbol{z}_{2}\right]$,
and similarly $\langle\ldots\rangle_{\mathcal{{W}^{\prime}}}$ over
noises $\boldsymbol{z}_{1}^{\prime},\boldsymbol{z}_{2}^{\prime}$
drawn from the transformed distribution $\mathcal{W}^{\prime}\left[\boldsymbol{z}_{1}^{\prime},\boldsymbol{z}_{2}^{\prime}\right]$,
with $\tilde{\boldsymbol{\rho}}(t)$ being evolved using the $\boldsymbol{z}_{1}^{\prime},\boldsymbol{z}_{2}^{\prime}$
noises. This technique is well understood in the context of stochastic
Schrodinger equations \citep{diosi1998non,strunz1999open,tanimura2006stochastic}.

Performing a Girsanov transformation of the SLN Eq. \eqref{eq:trace_preserving},
we arrive at an alternative equation of motion (see Appendix \ref{sec:Girsanov-transformation}
for details) which we refer to as the guided SLN,
\begin{equation}
i\hbar\frac{d\boldsymbol{\rho}(t)}{dt}=\left[H(t),\boldsymbol{\rho}(t)\right]-\left(\eta(t)+\frac{i}{\hbar}\int_{0}^{t}dt^{\prime}K_{\eta\nu}(t-t^{\prime})\sigma(t^{\prime})\right)\left[\sigma_{z},\boldsymbol{\rho}(t)\right]-\frac{\hbar}{2}\nu(t)\left\{ \sigma_{z}-\sigma(t),\boldsymbol{\rho}(t)\right\} ,\label{eq:unravelled}
\end{equation}
noting that $\boldsymbol{\rho}(t)$ is evolved rather than $\tilde{\boldsymbol{\rho}}(t)$,
with $\sigma(t)$ being the guide spin of Eq. \eqref{eq:guide_spin-1}.
From Eq. \eqref{eq:rho_girsanov}, the physical density matrix is
then obtained by averaging over realisations of this new guided dynamics.

Another equivalent strategy is to start from the trace-violating Eq.
\eqref{eq:ESLE-method-1} and divide $\boldsymbol{\rho}$ by its trace
at each time step. When performing stochastic sampling, the trace
still needs to be taken into account according to Eq. \eqref{eq:transformed_distribution}.
This can be avoided as shown above by shifting the mean of the $\eta$
noise which leads to the same Eq. \eqref{eq:unravelled} but without
the guide term $\sigma(t)$ in the anti-commutator, 
\begin{equation}
i\hbar\frac{d\boldsymbol{\rho}(t)}{dt}=\left[H(t),\boldsymbol{\rho}(t)\right]-\left(\eta(t)+\frac{i}{\hbar}\int_{0}^{t}dt^{\prime}K_{\eta\nu}(t-t^{\prime})\sigma(t^{\prime})\right)\left[\sigma_{z},\boldsymbol{\rho}(t)\right]-\frac{\hbar}{2}\nu(t)\left\{ \sigma_{z},\boldsymbol{\rho}(t)\right\} .\label{eq:normalised}
\end{equation}
The physical density matrix is then obtained by the stochastic average
$\boldsymbol{\rho}^{ph}(t)=\langle\boldsymbol{\rho}(t)/\text{Tr}\left({\boldsymbol{\rho}(t)}\right)\rangle$,
and we refer to this equation of motion as the normalised SLN.

To summarise, three forms of SLN have been derived here:
\begin{itemize}
\item The \emph{original }SLN\emph{,} Eq. \eqref{eq:ESLE-method-1}, which
is not trace-preserving;
\item The \emph{guided }SLN\emph{,} Eq. \eqref{eq:unravelled}, which preserves
the trace via a Girsanov transformation;
\item The \emph{normalised }SLN, Eq. \eqref{eq:normalised}, where the trace
of the density matrix is explicitly normalised.
\end{itemize}
Alternatively, it is straightforward to derive all three (original,
guided and normalised) versions of the spin dynamics, Eqs. \eqref{eq:xspin}-\eqref{eq:trace_evolution1}.
For completeness, we give below their guided form, equivalent to Eq.
\eqref{eq:unravelled}: 
\begin{equation}
\hbar\frac{d\sigma_{x}(t)}{dt}=-\left[\epsilon(t)-2\hat{\eta}(t)\right]\sigma_{y}(t)-i\nu(t)\frac{\sigma_{x}(t)\sigma_{z}(t)}{\text{Tr}\left({\boldsymbol{\rho}(t)}\right)}\label{eq:xspin_unravelled}
\end{equation}
\begin{equation}
\hbar\frac{d\sigma_{y}(t)}{dt}=-\Delta\sigma_{z}(t)+\left[\epsilon(t)-2\hat{\eta}(t)\right]\sigma_{x}(t)-i\nu(t)\frac{\sigma_{y}(t)\sigma_{z}(t)}{\text{Tr}\left({\boldsymbol{\rho}(t)}\right)}\label{eq:yspin_unravelled}
\end{equation}
\begin{equation}
\hbar\frac{d\sigma_{z}(t)}{dt}=\Delta\sigma_{y}+i\nu(t)\,\text{Tr}\left({\boldsymbol{\rho}(t)}\right)-i\nu(t)\frac{\sigma_{z}^{2}(t)}{\text{Tr}\left({\boldsymbol{\rho}(t)}\right)}\label{eq:zspin_unravelled}
\end{equation}
where $\text{Tr}(\boldsymbol{\rho}(t))$ is constant and $\hat{\eta}$ is simply the shifted $\eta$, 
\begin{equation}
\hat{\eta}(t)=\eta(t)+\frac{i}{\hbar}\int_{0}^{t}dt^{\prime}K_{\eta\nu}\left(t-t^{\prime}\right)\frac{\sigma_{z}(t)}{\text{Tr}\left({\boldsymbol{\rho}(t)}\right)},\label{eq:eta_hat}
\end{equation}
having written the guide spin $\sigma(t)$ in the form given by Eq.
\eqref{eq:guide_spin-1}. As before, the time-dependent spins here
represent quantum averages over a single stochastic density matrix
$\sigma_{i}(t)=\text{Tr}{\left(\sigma_{i}\boldsymbol{\rho}(t)\right)}$.

The same transformation has also recently been applied to density
matrices starting in partitioned or pure states and evolved via the
SLN \citep{stockburger2001non,stockburger2002exact,stockburger2004simulating},
though the reasoning was slightly different, thermalisation was not
included and no numerical results were shown. The authors started
from the original SLN and applied the transformation 
\[
\tilde{\boldsymbol{\rho}}(t)=\boldsymbol{\rho}(t)\,\exp\left\{ \frac{i}{\hbar}\int_{0}^{t}dt^{\prime}\nu\left(t^{\prime}\right)\gamma\left(t^{\prime}\right)\right\} 
\]
with $\gamma(t)$ being an unknown function. $\gamma(t)$ was later
chosen to enforce trace-preserving dynamics, leading to the obvious
choice $\gamma(t)=\sigma(t)$ and the equation identical to Eq. \eqref{eq:unravelled}.
Following the same steps as above, the exponential factor in the sampling
procedure is removed to arrive at the simple averaging of the trajectories.

\section{Stochastic Differential Equations\label{sec:Stochastic-Differential-Equation}}

It is well known that Langevin equations are ill-defined when expressed
as differential equations due to the white noise being everywhere
discontinuous \citep{risken1996fokker,kloeden2013numerical,platen1999introduction,gardiner1985handbook,gardiner2009stochastic,sarkka2019applied}.
Instead, discretised integral equations involving the Wiener process
increment are used to bring them into a well defined form. The standard
result for a set of coupled SDEs of a vector of functions $\boldsymbol{\rho}_{h}=\left\{ \rho_{h}^{k}\right\} $
is 
\begin{equation}
d\rho_{h}^{k}=a^{k}(t_{h},\boldsymbol{\rho}_{h})dt+\sum_{j}B^{kj}(t_{h},\boldsymbol{\rho}_{h})dW_{h}^{j},\label{eq:coupled_manynoise_index}
\end{equation}
where $\boldsymbol{a}\left(t_{h},\boldsymbol{\rho}_{h}\right)$ is
the deterministic (so-called drift) component of the dynamics and
the index $h$ is associated with the discrete proper time $t_{h}=h\,dt$.
$\boldsymbol{B}(t_{h},\boldsymbol{\rho}_{h})=\{B^{kj}(t_{h},\boldsymbol{\rho}_{h})\}$
is a matrix whose rows $\boldsymbol{b}^{k}(t_{h},\boldsymbol{\rho}_{h})$
are vectors associated with each $\boldsymbol{\rho}_{h}$, and $dW_{h}^{j}=W_{h+1}^{j}-W_{h}^{j}$
is the Wiener increment, where $W_{h}^{j}=\int_{0}^{t_{h}}dt^{\prime}x_{j}\left(t^{\prime}\right)$,
with $x$ being a white noise. This is just a first-order Taylor expansion
known as the Euler-Maruyama approximation or the Cauchy-Euler method
\citep{gardiner1985handbook}, for which the deterministic and stochastic
Taylor expansions are the same.

For a higher order scheme, additional terms that do not appear in
the deterministic Taylor expansion arise from the application of stochastic
calculus in either Stratonovich or Itô form \citep{platen1999introduction}.
For example, the second order Itô scheme, known as the Milstein scheme,
reads 
\begin{equation}
\rho_{h+1}^{k}=\rho_{h}^{k}+a^{k}(t_{h},\boldsymbol{\rho}_{h})dt+\sum_{j}B^{kj}(t_{h},\boldsymbol{\rho}_{h})dW_{h}^{j}+\sum_{l}\sum_{j_{1},j_{2}}B^{lj_{1}}(t_{h},\boldsymbol{\rho}_{h})\frac{B^{kj_{2}}(t_{h},\boldsymbol{\rho}_{h})}{\partial\rho_{h}^{l}}I_{j_{1},j_{2}}\label{eq:milstein}
\end{equation}
where 
\begin{equation}
I_{j_{1},j_{2}}=\int_{t_{h}}^{t_{h+1}}\int_{t_{h}}^{t_{h+1}}dW_{h}^{j_{1}}dW_{h}^{j_{2}}\label{eq:milstein_integral}
\end{equation}
is the Wiener integral. The solutions to these integrals grow in complexity
as the number of noises and/or the system size increases, though general
solutions are known \citep{kloeden2013numerical}. In addition, the
normal rules of calculus do not apply in Itô calculus but do for Stratonovich,
at the cost of introducing a correction which modifies the deterministic
drift \citep{gardiner1985handbook}. For the purposes of this work,
where many noises are necessary, Stratonovich calculus is more computationally
efficient with easier implementation and hence this interpretation
will be used. The dynamics still has the same form as Eq. \eqref{eq:coupled_manynoise_index},
but the drift $a^{k}(t_{h},\boldsymbol{\rho}_{h})$ is replaced by
the modified drift 
\begin{equation}
\tilde{a}^{k}(t_{h},\boldsymbol{\rho}_{h})=a^{k}(t_{h},\boldsymbol{\rho}_{h})-\frac{1}{2}\sum_{lj}B^{lj}(t_{h},\boldsymbol{\rho}_{h})\frac{\partial B^{kj}(t_{h},\boldsymbol{\rho}_{h})}{\partial\rho_{h}^{l}}.\label{eq:modified_drift}
\end{equation}

Since Stratonovich SDEs obey the rules of ordinary calculus, a family
of Runge-Kutta numerical methods can be developed. We shall use a
Heun scheme \citep{tzitzili2015numerical} with strong order convergence
of 1.0 compared to only 0.5 for the naive Euler-Maruyama approximation
\citep{platen1999introduction}, making it the same as the second
order Itô-Milstein scheme \citep{10.2307/2156972}. The Heun scheme
uses an intermediary prediction step to calculate a supporting value
$\hat{\boldsymbol{\rho}}_{h+1}$ which improves on an initial guess,
so that the next time step prediction becomes
\begin{equation}
\rho_{h+1}^{k}=\rho_{h}^{k}+\frac{1}{2}\left(\tilde{a}^{k}\left(t_{h},\boldsymbol{\rho}_{h}\right)+\tilde{a}^{k}\left(t_{h},\hat{\boldsymbol{\rho}}_{h+1}\right)\right)dt+\frac{1}{2}\sum_{j}\left(B^{kj}(t_{h},\boldsymbol{\rho}_{h})+B^{kj}(t_{h},\hat{\boldsymbol{\rho}}_{h+1})\right)dW_{h}^{j},\label{eq:huen}
\end{equation}
where the supporting value $\hat{\boldsymbol{\rho}}_{h+1}$ is obtained
via an Euler-Maruyama integrator with the Stratonovich correction,
\begin{equation}
\hat{\rho}_{h+1}^{k}=\rho_{h}^{k}+\tilde{a}^{k}(t_{h},\boldsymbol{\rho}_{h})dt+\sum_{j}B^{kj}(t_{h},\boldsymbol{\rho}_{h})dW_{h}^{j}.\label{eq:supporting_value}
\end{equation}

The derivation of the final Stratonovich corrections in imaginary
and real time are provided in Appendix \ref{sec:Stratonovich-Correction-for}.
We give there the explicit form of the Heun scheme of Eq. \eqref{eq:huen}
for the spin-boson model in terms of the components of the density
matrix, as well as for mean $x$, $y$ and $z$ spins. Note that there
is no correction for the trace.

The final procedure for the numerical solution of the ESLN is as follows:
\begin{enumerate}
\item Generate the appropriate filtering kernels $G\left(t\right)$, Eqs.
\eqref{eq:filter-eta-eta}-\eqref{eq:filter-eta-mu}, from the model
specific physical kernels $K\left(t\right)$, Eqs. \eqref{eq:kernel_etaeta}-\eqref{eq:kernel_mumu},
via application of the discrete Fourier transform and its inverse.
\item For each new realisation of the stochastic density matrix, generate
a set of orthogonal noise components $\eta_{\eta}$, $\eta_{\nu}$,
$\eta_{\mu},$ $\nu_{\eta}$, $\mu_{\eta}$ and $\mu_{\mu}$.
\item Rescale the $\eta_{\nu}$, $\nu_{\eta}$ and $\eta_{\mu},$ $\mu_{\eta}$
noises as required, as detailed in Section \eqref{sec:variance_reduction}.
\item Initialise the pre-thermalised density matrix in the state $\overline{\boldsymbol{\rho}}\left(\tau=0\right)=\mathbb{{I}}$
before evolving in imaginary time for $\tau\in[0,\beta\hbar]$, using
the Stratonovich modified drift as detailed in Appendix \ref{sec:Stratonovich-Correction-for},
Eq. \eqref{eq:thermalisation_modified_drift}.
\item Initialise the real time stochastic density matrix using the final
value from the imaginary time evolution, $\boldsymbol{\rho}\left(t_{0}\right)=\overline{\boldsymbol{\rho}}\left(\beta\hbar\right)$.
Evolve it in real time with the Stratonovich modified drift, Eq. \eqref{eq:real_time_modified_drift}
in Appendix \ref{sec:Stratonovich-Correction-for}. If desired, one
of the trace preserving variants of Eqs. \eqref{eq:unravelled} and
\eqref{eq:normalised} may be used. Alternatively, spin dynamics given
by Eqs. \eqref{eq:xspin}-\eqref{eq:trace_evolution1} can be used
instead, with the corresponding Stratonovich corrections, Eqs. \eqref{eq:Sx}-\eqref{eq:SzTr}.
\item Repeat the simulation (points 4-5) as many times as required, before
taking the ensemble average over the realisations of the density matrix,
then divide by the value of the trace of the ensemble average after
thermalisation $\text{Tr}\left(\langle\boldsymbol{\rho}(t_{0})\rangle\right)$
to obtain the physical density matrix.
\end{enumerate}

\section{Results\label{sec:Results}}

\subsection{Noise and Convergence\label{subsec:noise_and_convergence}}

\begin{figure}[h!]
\includegraphics[width=.32\linewidth]{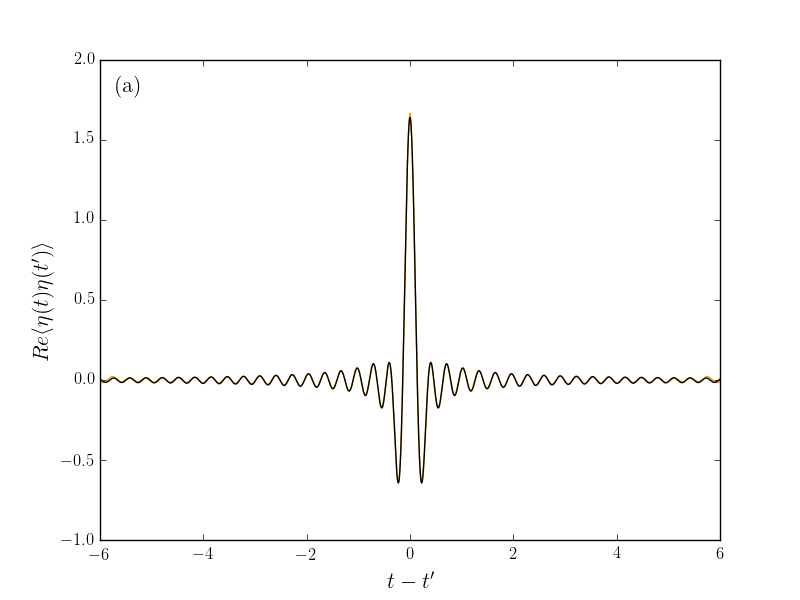}
\includegraphics[width=.32\linewidth]{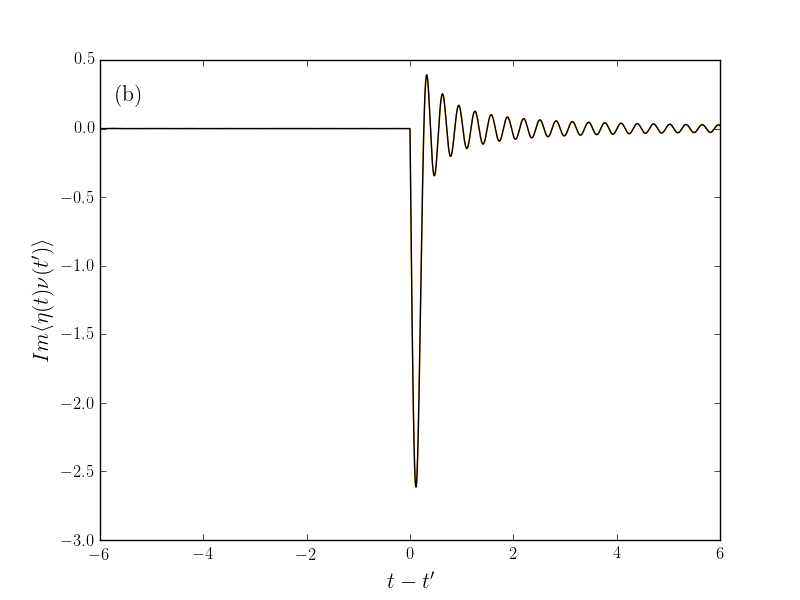}
\includegraphics[width=.32\linewidth]{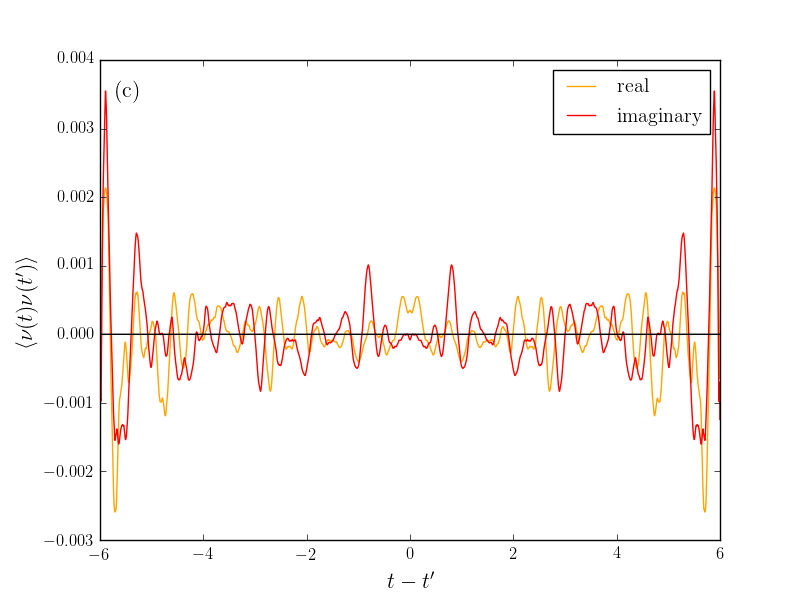}
\includegraphics[width=.32\linewidth]{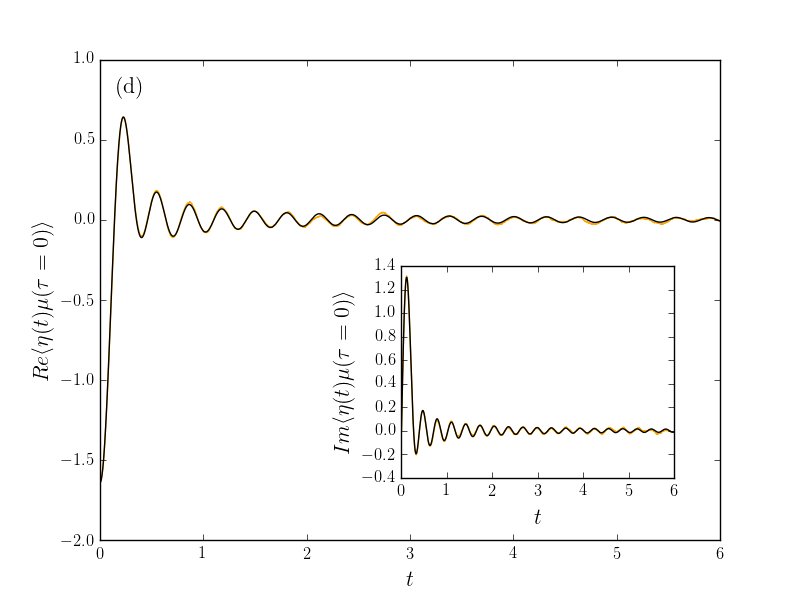}
\includegraphics[width=.32\linewidth]{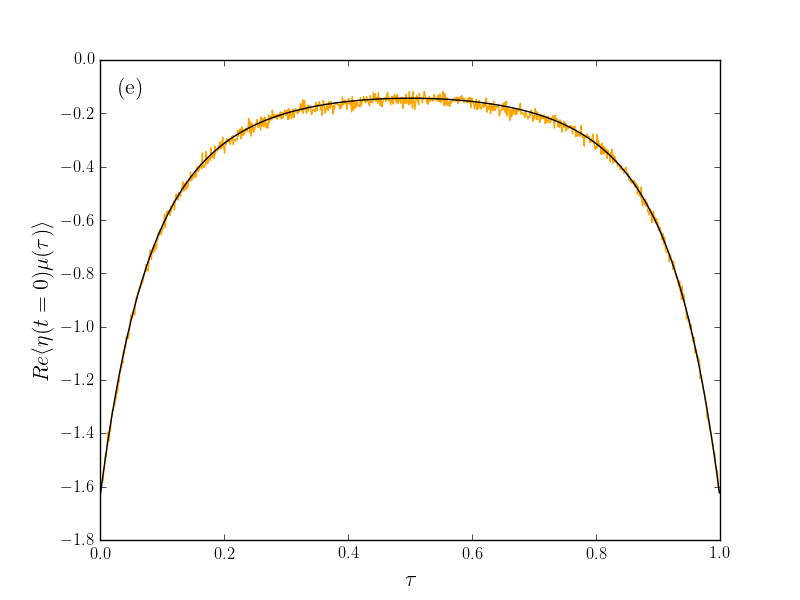}
\includegraphics[width=.32\linewidth]{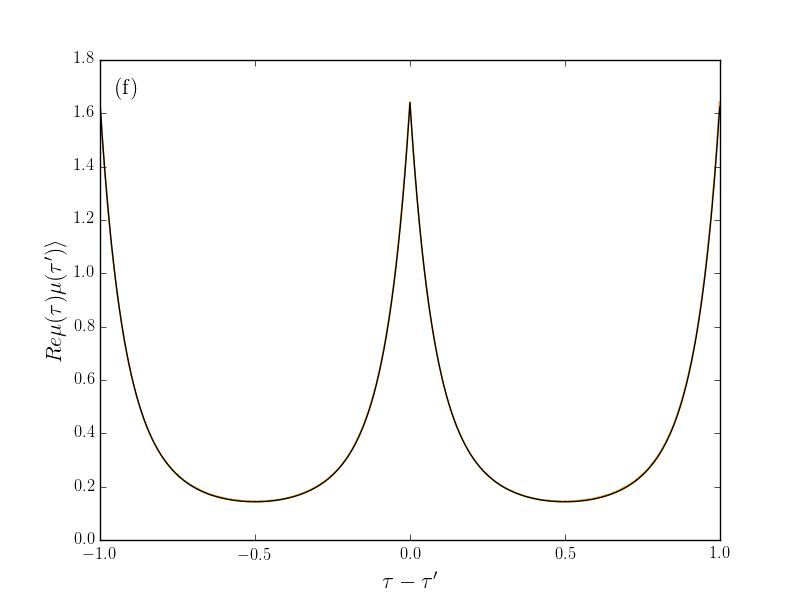}
\caption{Correlation functions of the noises, where the black line is the appropriate
kernel given in Eqs. \eqref{eq:kernel_etaeta}-\eqref{eq:zero_correlations}
and the orange line is the numerical correlation computed for 1 million
realisations with $\beta\hbar=1$, $t_{max}=6$, $dt=d\tau=10^{-3}$
and $\omega_{c}=20$. If only the black curve is visible, the orange
curve lies exactly underneath. All other correlations (not shown)
are zero to within 0.001. (a) The $\eta-\eta$ auto-correlation. (b)
The $\eta-\nu$ correlation. (c) The $\nu-\nu$ auto-correlation,
which is zero as required (within the adopted precision). (d) The
real part of the $\eta-\mu$ correlation when $\tau=0$, with the
imaginary part given in the inset. (e) The $\eta-\mu$ correlation
when $t=0$. (f) The $\mu-\mu$ auto-correlation. Optimal scaling
of $r_{\nu\eta}=0.5$ with $r_{\mu\eta}=1$ has been used in all cases.
\label{fig:correlations}}
\end{figure}

Using the noise generation procedure detailed in Section \ref{sec:noise_generation}
where the noise components are generated in Fourier space before taking
the inverse Fourier transform, it is found that the required correlation
functions (Figure \ref{fig:correlations}) are satisfied and converge
well. The cross-correlated noise $\eta_{\mu}$ presents a computational
bottleneck in terms of simulation time, since Fourier methods cannot
be employed and weighted sums of white noise random numbers must be
computed directly. Choosing $\eta_{\mu}$ as a coloured noise and
$\mu_{\eta}$ as a white noise also reduces the rate at which the
cross-time correlation matrix converges with sample size, making the
cross-correlated noise generation doubly expensive \citep{noise-generation-Matt-Dan-2020}.
No alternative choice is known to us at the time of writing.

\begin{figure}[h!]
\centering{}\includegraphics[width=.6\linewidth]{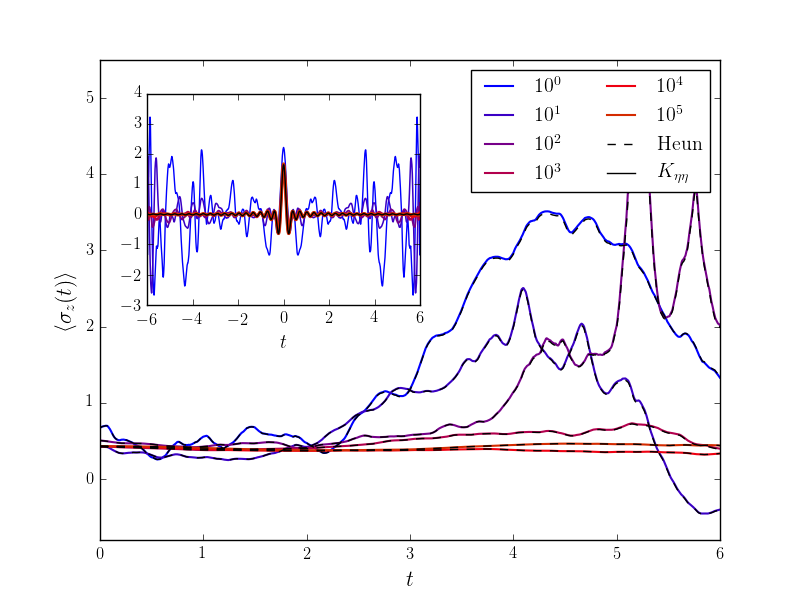}
\caption{Convergence of $\langle\sigma_{z}(t)\rangle$ for a constant Hamiltonian
with $\Delta=1$ and $\epsilon=-1$ for the Euler-Maruyama (solid
coloured lines) and Heun (dashed black lines) schemes for the same
sets of parameters, both performed for a range of sample sizes and
using the original SLN of Eq. \eqref{eq:ESLE-method-1}. The inset
shows the $\eta-\eta$ correlation for the same range of sample sizes,
as well as the corresponding physical kernel $K_{\eta\eta}(t-t')$
(black line). $\beta\hbar=1$, $t_{max}=6$, $dt=d\tau=10^{-3}$,
$\alpha=0.05$, $\omega_{c}=20$, $r_{\nu\eta}=0.5$ and $r_{\mu\eta}=1$.\label{fig:convergence}}
\end{figure}

Next we discuss the importance of the higher order numerical scheme
(Heun) considered in Section \eqref{sec:Stochastic-Differential-Equation}
(and derived in Appendix \ref{sec:Stratonovich-Correction-for}) in
solving the SDEs. To this end, we shall consider the real time dynamics
of $\langle\sigma_{z}(t)\rangle$ for a constant spin-boson Hamiltonian,
initialised in the proper thermal state. In Figure \ref{fig:convergence},
we compare the convergence properties of $\langle\sigma_{z}(t)\rangle$
for increasing sample size using both the Euler-Maruyama and Heun
discretisation schemes.

The expected result is for the spin to remain constant and equal to
the value obtained during thermalisation ($t=0$) during all real
times $t\geq0$; this behaviour is only evident for sufficiently large
sample size. It is clear that the error depends almost entirely on
the properties of the noises rather than inclusion of higher order
dynamical terms coming from stochastic calculus, since the results
obtained using the Heun scheme are indistinguishable from the Euler-Maruyama
scheme. This indicates that the convergence is solely statistical,
depending almost entirely on the sample size. However, in the special
case of weak coupling being simulated out to late times when the statistical
convergence is well controlled, the Heun scheme is necessary. If the
Euler-Maruyama scheme is used, the coherences ($x$ and $y$ spins)
oscillate within a exponentially growing envelope at late times, whereas
the Heun scheme reduces the time stepping error sufficiently to recover
decoherent dynamics. For this reason we use the Heun scheme for all
subsequent results, but note that future work should focus on optimising
the noise generation method for better convergence instead of improving
the discretisation scheme. The Stratonovich corrections of Eqs. \eqref{eq:Sx}
and \eqref{eq:Sy} are also used in all subsequent results for completeness,
though their effect is negligible. This is unsurprising since they
are of order $\mathcal{O}\left(dt\right)$. Finally, the $\eta-\eta$
correlation in the inset in Figure \ref{fig:convergence} uses the
same noises as the dynamics, averaged over the same number of runs,
emphasising the equivalence between the convergence of the noise correlations
with the convergence of the sample dynamics.

\begin{figure}[h!]
\centering{}\includegraphics[width=.6\linewidth]{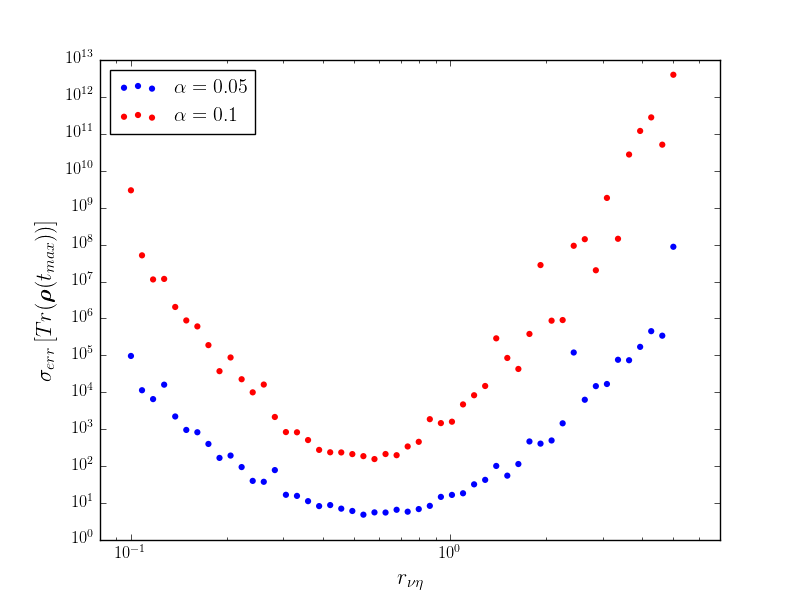}
\caption{The standard error of the mean $\text{Tr}\left(\boldsymbol{\boldsymbol{\rho}}(t_{max})\right)$
at its final time step for several values of the scaling factor $r_{\nu\eta}$.
For each scaling factor, 10 thousand runs for real time dynamics were
performed. $\beta\hbar=1$, $t_{max}=10$, $dt=d\tau=10^{-3}$, $\omega_{c}=20$
and $\epsilon=\Delta=0$. In this case, the minimising value of $r_{\nu\eta}$
is $\approx\frac{1}{2}$.\label{fig:error_mean_trace}}
\end{figure}

Though the correlation functions can be obtained for any choice of
scaling $r_{\mu\eta}$ and $r_{\nu\eta}$ introduced in Section \eqref{sec:variance_reduction},
choosing $r_{\nu\eta}$ to minimise the growth of the trace should
extend the time accessible by simulation. The effect of increasing
$r_{\nu\eta}$ from 0.1 to 5 on the standard error of the mean trace
$\text{Tr}{\left(\langle\boldsymbol{\rho}\left(t_{max}\right)\rangle\right)}$
for a sample of 10 thousand realisations is shown in Figure \ref{fig:error_mean_trace}
for two different environment coupling strengths $\alpha$. It is
tempting to take the limit where $r_{\nu\eta}$ becomes large and
$\nu\rightarrow0$ so that the dynamics becomes exactly trace preserving,
see Eq. \eqref{eq:trace_evolution1}, but such a choice would cause
$\eta_{\nu}$ to be very large, leading to poor convergence or instability.
We see that the order of magnitude of the error increases rapidly
beyond a narrow band of ratios for which it is at a minimum around
$r_{\nu\eta}\approx0.5$, evidence that the convergence of the system
dynamics is very sensitive to the properties of the noises even when
they satisfy the necessary correlation functions. In all subsequent
results, scaling of $r_{\nu\eta}=0.5$ and $r_{\mu\eta}=1.0$ are
used. While similar optimal scaling $r_{\mu\eta}$ could be chosen
to minimise the spread of initial values from thermalisation, the
variance is not significant and the simpler choice of $r_{\mu\eta}=1.0$
is sufficient. It is also clear that increasing the coupling strength
$\alpha$ makes the convergence worse as expected, since the noise
amplitudes scale like $\sqrt{{\alpha}}$. Unlike the noise amplitudes,
the variance within a sample grows non-linearly with $\alpha$ rather
than $\sim\sqrt{\alpha}$.

\subsection{Thermalisation}

\begin{figure}[h!]
\includegraphics[width=.49\linewidth]{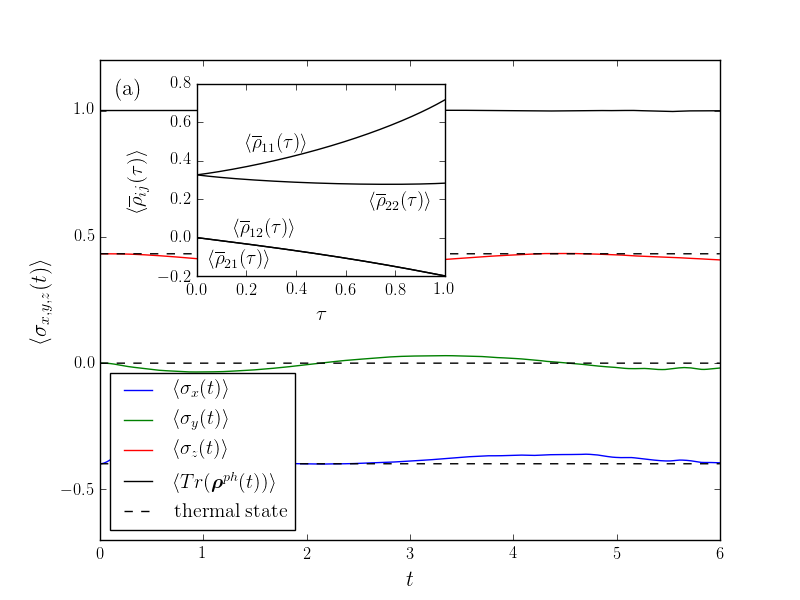}
\includegraphics[width=.49\linewidth]{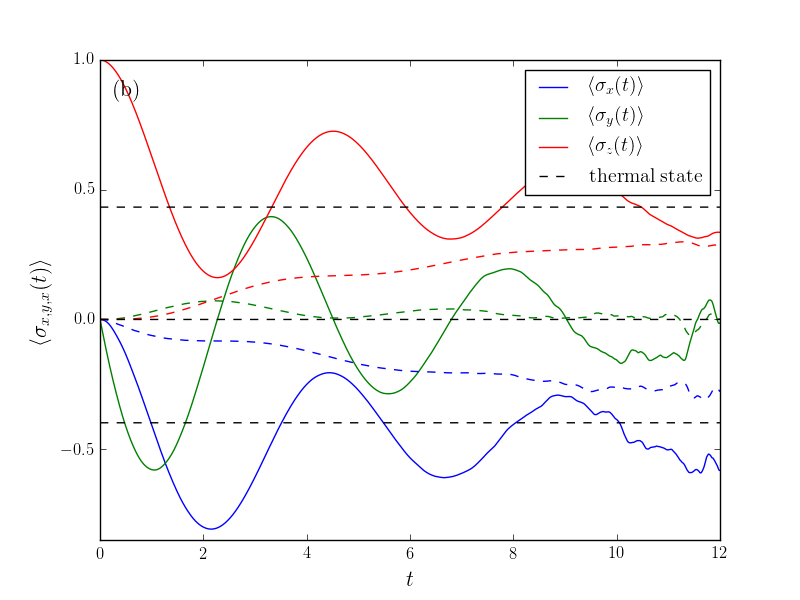}
\caption{Physical spins evolved by means of the original SLN, Eq. \eqref{eq:ESLE-method-1},
for different initial conditions using a constant Hamiltonian with
$\Delta=1$ and $\epsilon=-1$. Other simulation parameters are $\beta\hbar=1$,
$t_{max}$ as shown, $d\tau=dt=10^{-3}$, $\alpha=0.05$ and $\omega_{c}=20$.
Dashed black lines are the values of the thermalised spins obtained
from the end of imaginary time evolution. (a) Each realisation was
initialised in the canonical equilibrium state obtained from thermalisation
in imaginary time and averaged over 10 million runs. Inset: the elements
of the density matrix during imaginary time evolution. (b) Initially
decoupled from the environment and initialised out of equilibrium,
the spin components all decay towards the correct canonical equilibrium
state (black dashed lines) as obtained separately from thermalisation.
100 million realisations were used. Since there was no thermal preparation,
the $\eta$ noise has no $\eta_{\mu}$ component. Spins initialised
in the pure initial state $\sigma_{z}(0)=1$, $\sigma_{x}(0)=\sigma_{y}(0)=0$
are given by the solid coloured lines, while the zero initial state
$\sigma_{x}(0)=\sigma_{y}(0)=\sigma_{z}(0)=0$ spins are given by
dashed coloured lines.\label{fig:thermalisation_and_decay}}
\end{figure}

The ESLN is unique in its ability to simulate quantum dynamics exactly,
starting in the canonical equilibrium state with system-environment
entanglement arising from joint preparation. In Figure \ref{fig:thermalisation_and_decay}(a),
stationary state dynamics for the spin-boson system is shown using
the original SLN of Eq. \eqref{eq:ESLE-method-1}, with the open system
having been initialised in the thermal state via evolution in imaginary
time (Eq. \eqref{eq:thermalisation}). Small amplitude oscillations
around the equilibrium state are observed, most likely caused by variation
in the initial condition arising from the stochastic nature of thermalisation,
and vanishing as the sample size increases. For completeness, the
elements of the pre-thermalised density matrix $\langle\overline{\boldsymbol{\rho}}(\tau)\rangle$
are included in the inset, being evolved in $\tau$ from the initial
unitary state at $\tau=0$ to the thermal state at $\tau=\beta\hbar$.
This is the physical expectation obtained by the ensemble average
over many realisations of the environment noises, divided by the final
trace after averaging. The physical trace is divided by $\text{Tr}{\langle\overline{\boldsymbol{\rho}}(\beta\hbar)\rangle}$
to ensure that $\text{Tr}{\left(\boldsymbol{\rho}^{ph}(t_{0})\right)}=1$.

It is also necessary to check that the system decays to the correct
thermal state after being initially partitioned from the environment.
In Figure \ref{fig:thermalisation_and_decay}(b), the open system
was initialised in the pure state $\rho_{ij}(0)=\delta_{i1}\delta_{j1}$
(solid lines),. corresponding to $\sigma_{z}(0)=1$ and $\sigma_{x}(0)=\sigma_{y}(0)=0$.
In another simulation (dashed lines) the density matrix was initialised
in the half-half state $\rho_{11}=\rho_{22}=\frac{1}{2}$ and $\rho_{12}=\rho_{21}=0$,
which corresponds to the spin-zero state $\sigma_{x}(0)=\sigma_{y}(0)=\sigma_{z}(0)=0$.
In both cases, the coupling to the environment was switched on at
$t=0$ so that the system then begin to thermalise. Clearly, when
initialised in both the pure state $\sigma_{z}(0)=1$ (coloured solid
lines) and the $\sigma_{z}(0)=0$ state (coloured dashed lines), the
spins decay towards the thermal state as obtained from imaginary time
evolution (black lines) in the manner expected.

\subsection{Forms of the ESLN}

Since each realisation of the trace undergoes noisy growth within
an exponential envelope, Eq. \eqref{eq:trace_evolution1}, such that
the average trace converges poorly (see Figure \ref{fig:error_mean_trace}),
it may be desirable to use one of the trace preserving variants of
the ESLN: the guided ESLN of Eq. \eqref{eq:unravelled} or the normalised
ESLN of Eq. \eqref{eq:normalised}. In both cases, the physical trace
after the ensemble average should be constant. As for individual realisations,
in the case of guided dynamics, the trace is preserved exactly since
the guide spin forces the derivative of the trace to be zero. For
individual realisations of the normalised dynamics however, the trace
is not required to be constant and the ensemble average is taken over
$\boldsymbol{\rho}(t)/\text{Tr}\left({\boldsymbol{\rho}}(t)\right)$
rather than over $\boldsymbol{\rho}(t)$, forcing the physical trace
to be one.

\begin{figure}[h!]
\includegraphics[width=.49\linewidth]{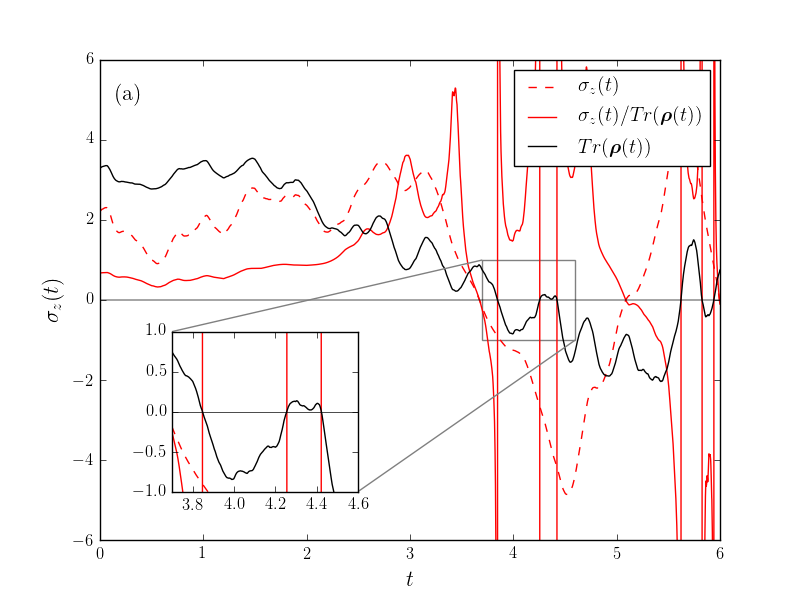}
\includegraphics[width=.49\linewidth]{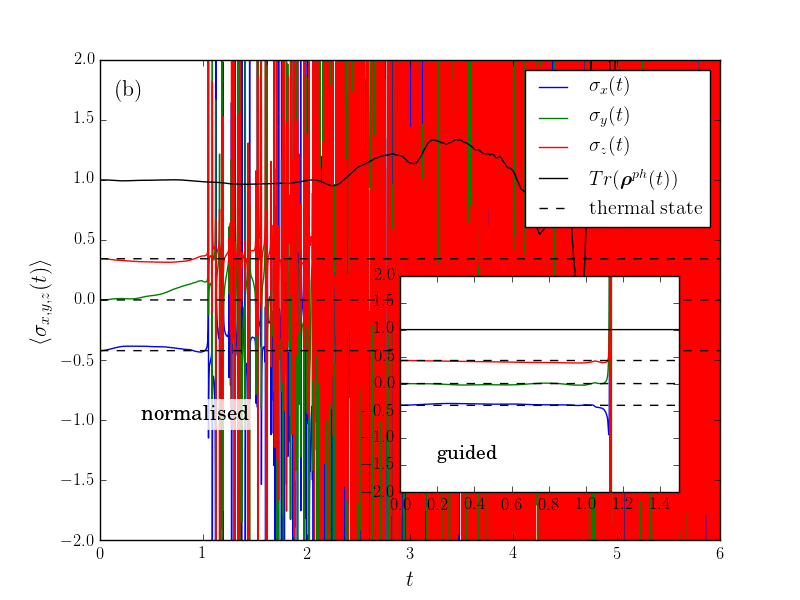}
\caption{(a) A single realisation of the normalised ESLN spin dynamics, where
$\sigma_{z}(t)$ is the $z-$spin, evolved by Eq. \eqref{eq:normalised}.
The inset highlights the behaviour when $\text{Tr}\left({\boldsymbol{\rho}(t)}\right)$
crosses zero. The magnitude of the spikes in $\sigma_{z}(t)/\text{Tr}{\left(\boldsymbol{\rho}(t)\right)}$
at these points reaches $\approx1300$. (b) 1 thousand realisations
of the normalised (main figure) and guided (inset) ESLN spin dynamics.
In all cases $\beta\hbar=1$, $t_{max}$ as shown, $dt=d\tau=10^{-3}$,
$\Delta=1$, $\epsilon=-1$, $\alpha=0.05$ and $\omega_{c}=20$ were
used. \label{fig:unravelled}}
\end{figure}

Figure \ref{fig:unravelled}(a) shows example dynamics for a single
realisation of the $z-$spin evolved using the guided ESLN of Eq.
\eqref{eq:unravelled}. In Figure \ref{fig:unravelled}(b), the spins
are evolved using the normalised ESLN of Eq. \eqref{eq:normalised},
and the guided ESLN for comparison and averaged over an ensemble of
1000 realisations. A single realisation of the $z-$spin for the original
ESLN of Eq. \eqref{eq:ESLE-method-1} can be found in Figure \ref{fig:convergence},
and an ensemble average in Figure \ref{fig:thermalisation_and_decay}(a);
the size of the ensemble average is not the same as in Figure \ref{fig:unravelled}(b),
but this does not affect the point being made here. In the case of
the normalised ESLN, the trace of a single spin trajectory is not
required to be constant or even positive at all times. Since the trace
is always initially positive, there are individual realisations where
the trace crosses zero and becomes negative. Since the ensemble average
is taken over $\boldsymbol{\rho}(t)/\text{Tr}\left({\boldsymbol{\rho}(t)}\right)$,
the physical density matrix and its observables will exhibit large
(infinite) spikes whenever $\text{Tr}\left({\boldsymbol{\rho}(t)}\right)=0$;
however, in practice it is unlikely that the trace would ever be exactly
zero so the spikes remain finite. Figure \ref{fig:unravelled}(a)
is an example of such a pathological trajectory. As a result, even
a small sample of 1 thousand realisations as in Figure \ref{fig:unravelled}(b)
accumulates many spikes, completely destroying the physical dynamics.
The averaged trace in Figure \eqref{fig:unravelled}(b) also fails
to be constant, since individual realisations of the trace are computed
directly (see black line in Figure \eqref{fig:unravelled}(a)) and
their averages are obtained in the normal way. The variation in the
average trace is thus an indication of undersampling only, whereas
the rapid fluctuation of the spins is largely independent of the sampling,
arising only from this division by (nearly) zero.

Individual realisations of the spins and trace evolved via the guided
ESLN are qualitatively similar to those evolved by the normalised
ESLN, with the exception that the guided trace is constant by definition;
it is not simulated directly but remains at its initial value $\text{Tr}\left({\overline{\boldsymbol{\rho}}(\beta\hbar)}\right)$.
This is true even when $\boldsymbol{\rho}(t)$ is simulated rather
than the spins and trace, in which case $\rho_{11}+\rho_{22}$ stays
constant to within $\pm10^{-13}$ of its initial value. However, the
guided ESLN includes a term containing the guide spin of Eq. \eqref{eq:guide_spin-1},
$\sigma(t)=\text{Tr}{\left(\sigma_{z}\boldsymbol{\rho}(t)\right)}/\text{Tr}{\left(\boldsymbol{\rho}(t)\right)}=\sigma_{z}(t)/\text{Tr}{\left(\boldsymbol{\rho}(t)\right)}$,
in which the $z-$spin is divided by the trace. This is just as pathological
as taking the ensemble average of $\boldsymbol{\rho}(t)/\text{Tr}{\left(\boldsymbol{\rho}(t)\right)}$
rather than $\boldsymbol{\rho}(t)$ in the normalised ESLN, since
the guide introduces the (possibly infinite) spikes directly into
the dynamics of individual trajectories. The system is usually unable
to recover, with individual realisations of the spins exceeding the
maximum allowed integer size of $2^{63}-1$. The ensemble average
similarly diverges, after which time the expectation values cease
to be physically meaningful. An example for a sample of 1 thousand
realisations is shown in the inset in Figure \ref{fig:unravelled}(b).
For both the guided ESLN and the normalised ESLN in Figure \ref{fig:unravelled},
the breakdown occurs at $t\sim1.1$. This feature is intrinsic to
the equations of motion themselves, and cannot be removed using a
larger sample since the probability of including a trajectory where
a spike occurs at $t\leq1.1$ increases with sample size.

Such behaviour occurs regardless of whether the equations of motion
for the density matrix or the spins are used, and does not appear
to depend on the parameters chosen in any meaningful way. While Eqs.
\eqref{eq:unravelled} and \eqref{eq:normalised} with their corresponding
ensemble averages analytically describe the correct physical dynamics,
the averages appear to be valid only in the limit that the sample
size is infinite. That is, for the analytic path integral of the distribution
$\mathcal{W}$ over the noise variables $\boldsymbol{z}_{1},\boldsymbol{z}_{2}$,
rather than a statistical average as is practically obtained for which
the results are pathological. Thus improvements in convergence to
address the growth of the trace must be obtained via other methods,
such as exploiting or even optimising the generation of the driving
noises \citep{noise-generation-Matt-Dan-2020}.

Concluding, both trace-conserving ESLN variants result in a pathological
behaviour in the dynamics that in practice cannot be cured by increasing
the sample size. Hence, in the following, only the original SLN, Eq.
\eqref{eq:ESLE-method-1}, is used.

\subsection{Landau-Zener Sweep}

\subsubsection{Modified Limit for Finite Temperature Coupling}

\begin{figure}[h!]
\includegraphics[width=.49\linewidth]{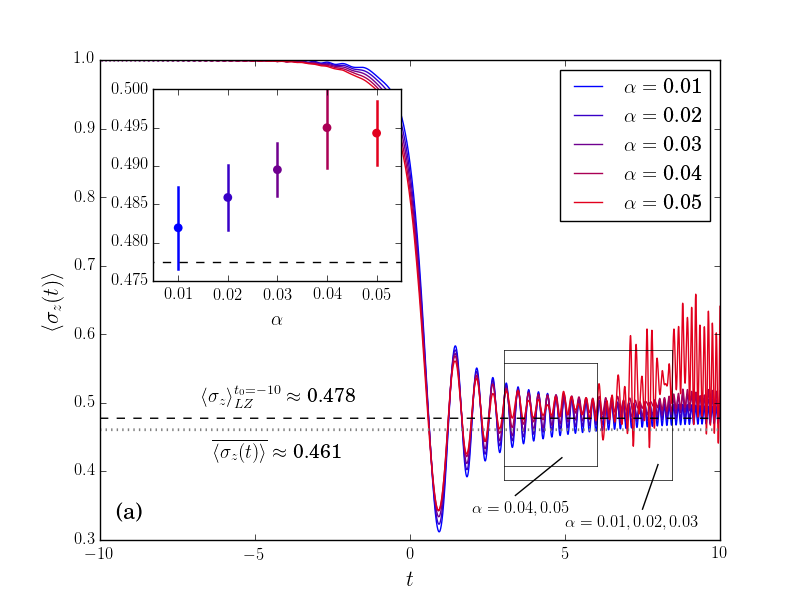}
\includegraphics[width=.49\linewidth]{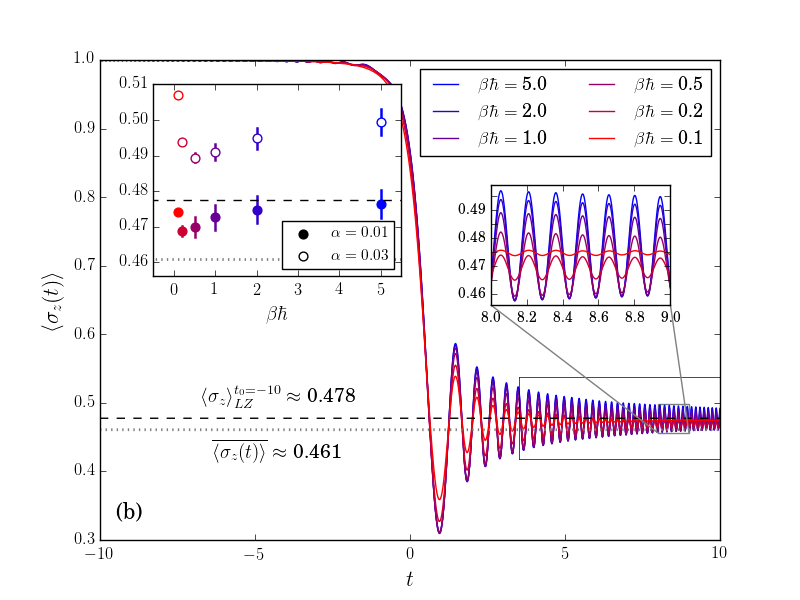}
\caption{Evolution of the physical $z-$spin $\langle\sigma_{z}(t)\rangle$
under LZ driving $\epsilon(t)=\kappa t$ with $\kappa=5$. The black
dotted and dashed lines are the original and modified (for $t_{0}=-10$)
LZ limits, respectively. (a) Dynamics for a range of coupling strengths
$\alpha\in[0.01,0.05]$ colour coded from blue to red with increasing
$\alpha$, all with the same temperature $\beta\hbar=1$, are shown.
The inset shows the observed asymptotic value for each coupling using
the same colours, obtained using 12 equally temporally spaced independent
estimates of the mean for $\alpha=0.01,0.02,0.03$ over 1 million
realisations and 6 equally spaced independent estimates of the mean
for $\alpha=0.04,0.05$ over 10 million realisations, taken over the
regions indicated by the labelled boxes. (b) Dynamics for a range
of inverse temperatures $\beta\hbar\in[0.1,5.0]$, colour coded from
blue to red with increasing temperature. As before, the inset on the
left shows the observed asymptotic value of the results using the
same colours, obtained using 20 independent estimates of the mean
over the boxed region. The solid circles are for the data shown in
the main figure with $\alpha=0.01$, while the empty circles are for
stronger coupling of $\alpha=0.03$ that remained well converged throughout
the simulation. The inset on the right shows the detailed dynamics
for times $8\protect\leq t\protect\leq9$ (see text). All other simulation
parameters are $dt=d\tau=10^{-3},$ $\Delta=1$, and $\omega_{c}=20$.\label{fig:LandauZener}}
\end{figure}

In Figure \ref{fig:LandauZener}, the spin-boson system is linearly
driven from negative to positive $\epsilon$ by a Landau-Zener (LZ)
sweep for a range of inverse temperatures $\beta\in[0.1,5.0]$ (panel
(a)) and environment coupling strengths $\alpha\in[0.01,0.05]$ (panel
(b)). The analytic LZ limit of Eq. \eqref{eq:LZlimit} is valid for
a spin which was initialised at zero temperature in its ground state
in the infinite past, $\sigma_{z}(-\infty)=1$, with all other spins
being zero. This limit describes the asymptotic state as $t\rightarrow\infty$
and while it was originally derived for an isolated spin \citep{zener1932non},
the result is valid for a zero temperature dissipative spin as well
\citep{wubs2006gauging,saito2007dissipative,rojo2010matrix,orth2013nonperturbative}
so is often used as a numerical test for approximate methods \citep{stockburger2002exact,stockburger2004simulating,saito2007dissipative,nalbach2009landau,orth2008dissipative,orth2013nonperturbative}.

\begin{figure}[h!]
\includegraphics[width=.6\linewidth]{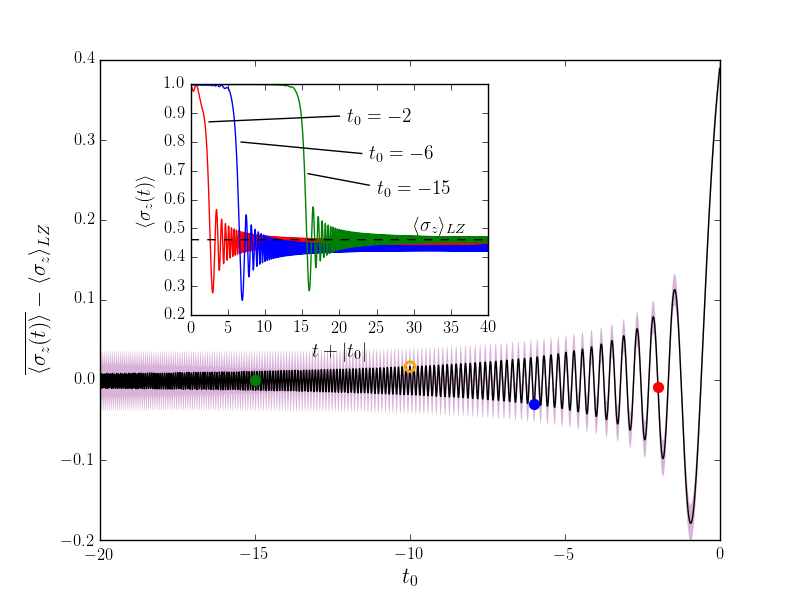}
\caption{The deviation of $\sigma_{z}(t)$ from the exact LZ limit for an isolated
spin (no bath, $\alpha=0$) with initial preparation $\sigma_{z}(t_{0})=1$.
The error on the average is the filled pink area, with the average
taken over the time period from the first maximum after the minimum
in the inset to the end of the simulation. The solid points are coloured
to correspond to the used $t_{0}$ values for the evolution examples
given in the inset. $t_{0}=-10$ is also highlighted (orange circle)
since this is the $t_{0}$ used in subsequent results. $dt=d\tau=10^{-3}$,
$\Delta=1$, $\epsilon(t)=\kappa t$ with $\kappa=5$. \label{fig:calibration}}
\end{figure}

In practice, the spin is initialised with $\sigma_{z}(t_{0})=1$ at
some finite time in the past $t_{0}<0$ instead, rather than when
$t_{0}\rightarrow-\infty$. This causes the late time dynamics to
approach a slightly different limit $\langle\sigma_{z}\rangle_{LZ}^{t_{0}}$
that deviates from the asymptotic limit, approaching $\langle\sigma_{z}\rangle_{LZ}$
only as $t_{0}\rightarrow-\infty$. This can clearly be seen in Figure
\ref{fig:calibration} where the deviation from the analytical limit
is calculated for the isolated system (no coupling to the bath, $\alpha=0$)
for many values of $t_{0}$. Thus Figure \ref{fig:calibration} acts
as a form of approximate calibration of the simulations with the bath
coupling turned on, allowing us to modify the LZ limit using the value
obtained for the isolated system to account for the finiteness of
$t_{0}$. Note however that this calibration alone is not sufficient
to fully correct the limit for finite $\beta$ and non-zero $\alpha$,
as Figure \ref{fig:calibration} was obtained for the closed system
with no bath rather than for an open system at zero temperature.

Using Figure \ref{fig:calibration}, we find that the deviation from
the modified limit $\langle\sigma_{z}\rangle_{LZ}^{t_{0}}$ (shown
by the dashed line) for a system coupled to a finite temperature bath
is larger for stronger coupling. This can been seen in Figure \ref{fig:LandauZener}(a).
It is apparent that for the largest coupling $\alpha=0.05$, the required
ensemble size becomes larger than the 10 million realisations used
here, which for the reasonably long simulation time $-10\leq t\leq10$
takes $\sim$15 hours on 360 CPUs compared to only $\sim1$ hour for
1 million realisations. As such, a smaller box has to be taken for
higher coupling when calculating the mean $z-$spin as an estimate
of the observed asymptote, see Figure \ref{fig:calibration}(a). This
poor convergence may explain the otherwise anomalous mean value for
$\alpha=0.05$ in the inset which moves towards the shifted limit
rather than away from it.

If the modified LZ limit $\langle\sigma_{z}\rangle_{LZ}^{t_{0}}$
for $t_{0}=-10$ had not been used, the observed asymptotes would
never approach the original LZ limit, not even in the $\alpha\rightarrow0$
limit. However, by using the modified limit we recover the expected
asymptotic dynamics for small $\alpha$ while stronger coupling forces
the $z-$spin away from the limit. This can be understood in terms
of the renormalised tunneling matrix element \citep{leggett1987dynamics},
\[
\Delta_{r}=\Delta\left(\frac{\Delta}{\omega_{c}}\right)^{\frac{\alpha}{1-\alpha}},
\]
which decreases with the coupling strength. After $t=0$, the $\sigma_z=-1$ 
state becomes the lower energy state with thermal fluctuations and tunnelling
contributing to the likelihood of a transition. Since $\Delta_r$ decreases with $\alpha$, 
the system is less likely to tunnel from $\sigma_{z}=+1$ to $-1$ for
larger $\alpha$, causing the observed increase in $\langle\sigma_{z}(t\lesssim t_{max})\rangle$.

In Figure \ref{fig:LandauZener}(b) we examine the behaviour of the
limit for a range of inverse temperatures $\beta\hbar\in[0.1,5.0]$,
and again find that the modified limit is required to observe the
expected asymptotic results; the original limit is missed altogether.
As the temperature is decreased, the observed asymptote tends towards
the modified limit as expected, with strange behaviour for high temperatures
(see inset, discussed below). Consistent with Figure \ref{fig:LandauZener}(a),
increasing the coupling to $\alpha=0.03$ (empty circles) from $\alpha=0.01$
(solid circles) in the inset has the effect of lifting the observed
asymptote, though the exact scaling of this shift for different $(\alpha,\beta)$
pairs has not been investigated as it is not of interest to us here.

For medium to high temperatures $0.5\leq\beta\hbar<2$, the asymptotic
$z-$spin decreases. This is as expected, since thermal fluctuations
in the bath serve to destroy coherence, with the mean of all the spin
components being zero in the high temperature limit. Strangely, for
very high temperatures $\beta\hbar<0.5$, the $z-$spin increases
towards the modified limit before surpassing it altogether. This is
not caused by poor statistical convergence, as is shown in the magnified
inset between $t=8$ and $9$ where the position of the curves clearly
increases for the two hottest temperatures. We suggest that this rapid
increase in the observed asymptote for higher temperatures occurs
as the energy scale of thermal fluctuations in the bath approaches
the typical energy separation between the two states at the end of
the simulation, $\epsilon(t_{max})$, providing enough energy for
the system to jump into the higher energy state. This is not a true
asymptotic effect, but a transient effect in the window $0<t\lesssim t_{max}$
that should vanish as $t\rightarrow\infty$. The dimensionless energy
ratio $q$ between the thermal energy scale of the bath, $k_{B}T$,
and the energy separation between the states, $\hbar\epsilon\left(t_{max}\right)$
(setting $\hbar=k_{B}=1$),

\[
q=\frac{k_{B}T}{\hbar\epsilon(t_{max})}=\frac{1/\beta}{\epsilon(t_{max})},
\]
will be of order 1 when thermal fluctuations are large enough to overcome
the finite bias within the simulation window. For the hottest temperature
in Figure \ref{fig:LandauZener}(b) $(\beta=0.1)$ the thermal energy
scale is $\sim10$ and the energy separation is $\sim50$ so that
$q=0.2$. While not of order one, an observable increase in the mean
spin would be expected, though the observed prominence of the high
temperature increase in the spin remains surprising.

\subsubsection{Thermalisation to recover the original limit}

It is possible to circumnavigate the need for a modified LZ limit
altogether by initialising the $z-$spin to be closer to the true
LZ spin at the actual finite (negative) $t_{0}$, rather than being
equal to one at $t_{0}$. The true LZ spin is initialised with $\sigma_{z}\left(-\infty\right)=1$
in the infinite past when the bias was infinitely large. It is obvious
that the change in spin acquired during its evolution from $-\infty$
up to the finite time $t_{0}$ would be different from $\sigma_{z}\left(t_{0}\right)=1$
which is commonly taken as the initial condition at the start time
of the simulation. In other words, the commonly simulated spin has
some \textquotedbl catching up\textquotedbl{} to do with respect to
the true LZ spin. As we shall demonstrate below, a more appropriate
initial state should recover the correct asymptotic dynamics without
needing to take the particular value of $t_{0}$ into account.

\begin{figure}[h!]
\includegraphics[width=.6\linewidth]{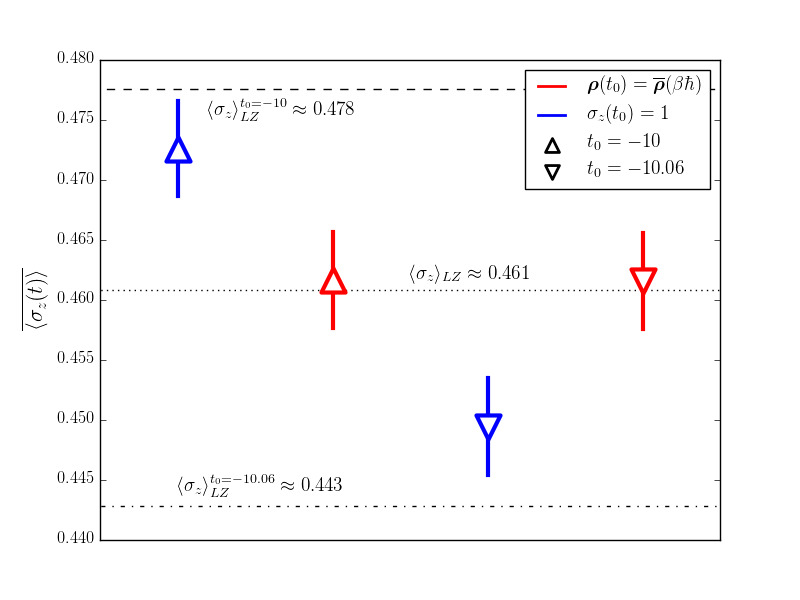}
\caption{Observed mean asymptotic $z-$spin when initialised with $\sigma_{z}(t_{0})=1$
and all other spin components zero (blue) and when the system is initially
thermalised in accordance with the initial value of the bias $\epsilon(t_{0})$
(red). Two different values of $t_{0}$ are shown, specifically chosen
so that one shifts the LZ limit upwards $(t_{0}=-10$) while the other
shifts it downwards $(t_{0}=-10.06)$, with the limits shown by dashed
black lines. The error on the mean was obtained using 15 independent
estimates of the mean in the region $3.5\protect\leq t\protect\leq10$.
Here $dt=d\tau=10^{-3},$ $\beta\hbar=1$, $\Delta=1$, $\epsilon(t)=\kappa t$
with $\kappa=5$, $\omega_{c}=20$, $\alpha=0.01$, averaged over
1 million realisations. \label{fig:thermalised}}
\end{figure}

For the dynamics over the period $-\infty<t\leq t_{0}$ the system
may be approximately thermalised; this should be exact in the adiabatic
limit of the LZ sweep rate $\kappa\rightarrow0$. Hence, one possible
state that we can choose at $t_{0}$ instead of the $t\rightarrow-\infty$
initial LZ value of $\sigma_{z}\left(t_{0}\right)=1$ would be the
\emph{equilibrium (thermalised) state }associated with $\epsilon(t_{0})$.
As long as $\left|\epsilon(t_{0})\right|$ is still much larger than
the other relevant energy scales of the system, this will be a good
approximation of the true LZ spin at $t_{0}$. The ESLN provides an
exact way of initialising the system in this equilibrium state by
the initial preparation in imaginary time, Eq. \eqref{eq:ESLE_thermalisation},
and as such the full ESLN represents an improvement on SLN methods
for modelling systems of this kind. Examples of the observed asymptote
of the $z-$spin are shown in Figure \ref{fig:thermalised} for both
initial conditions (thermalised and $\sigma_{z}\left(t_{0}\right)=1$)
to serve as a point of comparison.

One can see that if the modified limit is shifted up or down compared
to the original limit due to the finiteness of $t_{0}$, the thermalised
initial condition approaches the original limit very well, regardless
of the direction of the $t_{0}$ shift. Instead, it sits just above
$\langle\sigma_{z}\rangle_{LZ}$ as would be expected for a finite
simulation time rather than for one which runs to $t\rightarrow\infty$.
This cleanly demonstrates that thermalising the system with $\epsilon(t_{0})$
correctly accounts for the fact that the actual system is initialised
at the infinite past, and the original LZ limit is recovered. We emphasise
that this has not been achieved before.

\section{Discussion and Conclusions\label{sec:Discussion-and-Conclusions}}

In this paper we have demonstrated a successful implementation of
the Extended Stochastic Liouville-von Neumann equations (ESLN) computational
method for obtaining real time dynamics of the reduced density matrix
of an open quantum system coupled to a harmonic bath. This method
is exact and can be used for arbitrary open quantum systems at arbitrary
temperature and coupling strength, at least in principle, provided
that the coupling is linear in the bath coordinates. Unlike existing
SLN schemes where the system is initially decoupled from the bath,
in our method the combined system of the quantum system of interest
and the bath are fully thermalised together, with the coupling already
established. The main difference compared with SLN simulations is
that the density matrix has to be initially evolved in imaginary time
before the real time propagation for each sampling trajectory.

The utility of the method has been demonstrated on two simple systems
both based on the spin-boson Hamiltonian: (i) a case where the Hamiltonian
remains constant during real time evolution, and (ii) the Landau-Zener
(LZ) sweep that is a fully non-equilibrium evolution.

The first case was chosen to prove that our ESLN simulation can maintain
the thermalised state at any $t>0$ and also reach it asymptotically
if initialised in an arbitrary state. These simulations served as
a test bed for choosing the appropriate method of generating correlated
noises and establishing a computational scheme. We find that the noise
generation method proposed earlier \citep{mccaul2018driving} represents
the worst possible choice of the noises \citep{noise-generation-Matt-Dan-2020},
in the sense that the associated dynamics are highly unstable, restricting
simulation to only very short times. This presented something of a
paradox, in that the correlation functions were still fully satisfied
which is the only requirement of the theory on the noises. We provide
a modified scheme here, although other possibilities also exist \citep{noise-generation-Matt-Dan-2020}.
In addition, we have also concluded that trace-preserving variants
of the ESLN lead to pathological behaviour and so must be discarded,
leaving us with the original form of the SLN whose dynamics is not
unitary. As long as an appropriate noise generation scheme is used,
this turns out to be sufficient for the system to be well behaved.

Since the ESLN equations are stochastic in naturecare is required
in developing an appropriate numerical scheme. We find that in our
case the choice of the numerical scheme is essential if long time
simulations are needed, while for short timescales a regular Euler-Maruyama
discretisation is enough. For long time simulations the higher order
schemes originating from the Stratonovich stochastic calculus, such
as the Heun scheme, are advantageous.

Applications to the LZ model, in which the initial state is not thermalised,
also demonstrated that our method works well even in this rather complex
non-equilibrium situation. We showed that the LZ limit for a system
coupled to an environment differs from that of the isolated system,
but approaches it in the limit of zero environmental coupling and
temperature. We have also found that in actual simulations the asymptotic
limit is sensitive to the choice of the simulation time $t_{0}$ at
which the system is prepared. We find that this dependence can be
weakened substantially if the thermalised state corresponding to the
LZ Hamiltonian at the initial time is used instead of the correct
LZ initial state, demonstrating another utility of our method.

We hope that this work will stimulate further investigations of the
non-equilibrium dynamics of open quantum systems by means of the ESLN
method.

\appendix

\section*{Appendices}

\numberwithin{equation}{section}

\section{Girsanov transformation\label{sec:Girsanov-transformation}}

To ensure that the normalised $\tilde{\boldsymbol{\rho}}$ have the
correct physical ensemble average $\boldsymbol{\rho}^{ph}$, the transformation
$\mathcal{W}\rightarrow\mathcal{W}^{\prime}$ must take the form 
\begin{equation}
\mathcal{W}^{\prime}\left[\boldsymbol{z}_{1}^{\prime},\boldsymbol{z}_{2}^{\prime}\right]=\mathcal{W}[\boldsymbol{z}_{1},\boldsymbol{z}_{2}]\,\text{Tr}\left(\boldsymbol{\rho}(t)\right),\label{eq:transformed_distribution}
\end{equation}
as can be seen by substituting $\boldsymbol{\rho}(t)=\tilde{\boldsymbol{\rho}}(t)\,\text{Tr}\left(\boldsymbol{\rho}(t)\right)$
into Eq. \eqref{eq:rho_girsanov}. $\text{Tr}\left({\boldsymbol{\rho}(t)}\right)$
is easily obtained from Eq. \eqref{eq:trace_evolution1} as 
\begin{equation}
\text{Tr}\left(\boldsymbol{\rho}(t)\right)=\exp\left\{ \frac{i}{\hbar}\int_{0}^{t}dt'\,\nu\left(t^{\prime}\right)\sigma\left(t^{\prime}\right)\right\} .
\end{equation}
The task is now to remove this exponential factor from the average
by completing the square in $\mathcal{W}^{\prime}$ and identifying
transformed noises $\boldsymbol{z}^{\prime}$. Writing out Eq. \eqref{eq:transformed_distribution}
explicitly, this is 
\[
\mathcal{W}'\left[\boldsymbol{z}_{1}^{\prime},\boldsymbol{z}_{2}^{\prime}\right]=\mathcal{N}\exp\left\{ -\frac{1}{2}\left[\int_{0}^{t}dt^{\prime}\int_{0}^{t}dt^{\prime\prime}\boldsymbol{z}_{1}^{T}(t^{\prime})\boldsymbol{\Sigma}^{11}(t^{\prime}-t^{\prime\prime})\boldsymbol{z}_{1}(t^{\prime\prime})+2\int_{0}^{t}dt^{\prime}\int_{0}^{\beta\hbar}d\tau\boldsymbol{z}_{1}^{T}(t^{\prime})\boldsymbol{\Sigma}^{11}(t^{\prime},\tau)\boldsymbol{z}_{2}(\tau)\right.\right.
\]
\[
\left.\left.+\int_{0}^{\beta\hbar}d\tau\int_{0}^{\beta\hbar}d\tau^{\prime}\boldsymbol{z}_{2}^{T}(\tau^{\prime})\boldsymbol{\Sigma}^{22}(\tau-\tau^{\prime})\boldsymbol{z}_{2}(\tau^{\prime})\right]\right\} \exp\left\{ \frac{i}{\hbar}\int_{0}^{t}dt^{\prime}\mathbf{S}^{T}(t^{\prime})\boldsymbol{z}_{1}(t^{\prime})\right\} ,
\]
where vectors of noises $\boldsymbol{z}_{1}=\left(\eta\:\eta^{*}\:\nu\:\nu^{*}\right)^{T}$,
$\boldsymbol{z}_{2}=\left(\mu\:\mu^{*}\right)^{T}$ and $\boldsymbol{z}=\left(\boldsymbol{z}_{1}\:\boldsymbol{z}_{2}\right)^{T}$,
and the vector $\mathbf{\mathbf{S}}(t)=(0\:0\:\sigma(t)\:0)^{T}$
have been introduced. Note that the guide spin couples to the $\nu$
noise only in $\mathbf{S}(t)$. Making use of the fact that $\boldsymbol{\Sigma}$
is a symmetric matrix of time differences, and introducing its inverse
via 
\begin{equation}
\int_{0}^{t}dt^{\prime}\boldsymbol{\Sigma}^{-1}(s-t^{\prime})\,\boldsymbol{\Sigma}(t^{\prime}-t^{\prime\prime})=\delta(s-t^{\prime\prime}),
\end{equation}
the result of completing the square in symbolic notations is 
\[
-\frac{1}{2}\boldsymbol{z}^{T}\boldsymbol{\Sigma}\boldsymbol{z}+\frac{i}{\hbar}\mathbf{S}^{T}\boldsymbol{z}=A(t)-\frac{1}{2}\boldsymbol{z}^{\prime}\boldsymbol{\Sigma}\boldsymbol{z}^{\prime},
\]
where $A(t)$ is a function independent of $\boldsymbol{z}$ to be
absorbed into the normalisation of the physical density matrix $\mathbb{N}$,
and 
\begin{equation}
\boldsymbol{z}^{\prime}=\boldsymbol{z}-\frac{i}{\hbar}\boldsymbol{\Sigma}^{-1}\mathbf{S}
\end{equation}
are the transformed noises. These can be simplified by noting that
the vector $\mathbf{S}$ has just one $\nu$ non-zero component. Hence,
since $\nu$ is only correlated with $\eta$, only the $\eta$ noise
is modified, 
\begin{align}
 & \eta^{\prime}\left(t\right)=\eta\left(t\right)-\frac{i}{\hbar}\int_{0}^{t}dt^{\prime}K_{\eta\nu}\left(t-t^{\prime}\right)\sigma\left(t^{\prime}\right)\label{eq:eta_prime}\\
 & \nu^{\prime}\left(t\right)=\nu\left(t\right)\\
 & \mu^{\prime}\left(\tau\right)=\mu(\tau).
\end{align}

According to Eq. \eqref{eq:rho_girsanov}, the physical density matrix
is now 
\begin{equation}
\boldsymbol{\rho}^{ph}(t)=\int\mathcal{D}^{2}\left[\eta\right]\mathcal{D}^{2}\left[\nu\right]\mathcal{D}^{2}\left[\mu\right]\mathcal{W}^{\prime}\left[\eta^{\prime},\nu^{\prime},\mu^{\prime}\right]\tilde{\boldsymbol{\rho}}(t)\left[\eta,\nu,\mu\right],
\end{equation}
where the integrals are still performed over the original noises and
their complex conjugates. Since only $\eta$ is modified, the only
change of variables needed is $\eta\rightarrow\eta^{\prime}$ for
which the Jacobian $J=\left\lvert \frac{\delta\eta^{\prime}}{\delta\eta}\right\rvert $
contains the elements 
\begin{equation}
\frac{\delta\eta^{\prime}(t^{\prime})}{\delta\eta(t^{\prime\prime})}=\delta\left(t^{\prime},t^{\prime\prime}\right)-\frac{i}{\hbar}\int_{0}^{t}dsK_{\eta\nu}(t^{\prime}-s)\frac{\delta\sigma(s)}{\delta\eta(t^{\prime\prime})},\label{eq:transformation_Jacobian}
\end{equation}
where $K_{\eta\nu}\left(t^{\prime}-s\right)$ is a known correlation
function \eqref{eq:kernel_etanu}, independent of any particular realisation
of $\eta$, so does not need to be differentiated. It is also causal,
requiring that $t^{\prime}>s$, as is $\sigma(s)$, so $\delta\sigma(s)/\delta\eta(t^{\prime\prime})$
is only non-zero for $s>t^{\prime\prime}$. This bounds the integral
over $s$ from $t^{\prime\prime}$ to $t^{\prime}$ which corresponds
to a triangular matrix with zeros on the diagonal, and hence the integral
does not contribute to the determinant $J=\left\lvert \frac{\delta\eta^{\prime}}{\delta\eta}\right\rvert $.
Hence the Jacobian is simply equal to unity and applying the change
of variables $\eta\rightarrow\eta^{\prime}$ completes the transformation.
Since the transformed distribution has the same precision matrix $\boldsymbol{\Sigma}$
as the original distribution, the correlations for $\eta$ and $\eta^{\prime}$
have been preserved and the primes can be omitted. Replacing $\tilde{\boldsymbol{\rho}}$
with $\boldsymbol{\rho}$ for simplicity, we obtain the equation of
motion \eqref{eq:unravelled} given in the main text.

\section{Stratonovich Correction for the Spin-Boson model\label{sec:Stratonovich-Correction-for}}

\subsection{Real Time Propagation}

For the ESLN, it is convenient to rewrite the $2\times2$ density
matrix as a $4-$fold vector with elements $\rho_{h}^{k}$ (where
$k\in[1,4]$) with the original matrix elements ordered as 11, 12,
21, and 22. The dynamics is then split into one deterministic part
and two noisy parts associated with $\eta$ and $\nu$, 
\begin{equation}
d\rho_{h+1}^{k}=a^{k}(t_{h},\boldsymbol{\rho}_{h})dt+b_{\eta}^{k}(t_{h},\boldsymbol{\rho}_{h})\eta(t_{h})dt+b_{\nu}^{k}(t_{h},\boldsymbol{\rho}_{h})\nu(t_{h})dt\label{eq:ESLE_ab}
\end{equation}
where $a^{k}(t_{h},\boldsymbol{\rho}_{h})$, $b_{\eta}^{k}(t_{h},\boldsymbol{\rho}_{h})$
and $b_{\nu}^{k}(t_{h},\boldsymbol{\rho}_{h})$ in the right hand
side are elements of the vectors 
\begin{equation}
\boldsymbol{a}(t_{h},\boldsymbol{\rho}_{h})=-\frac{i}{\hbar}\left[H_{h},\boldsymbol{\rho}_{h}\right],\label{eq:a}
\end{equation}
\begin{equation}
\boldsymbol{b}_{\eta}(t_{h},\boldsymbol{\rho}_{h})=\frac{i}{\hbar}\left[\sigma_{z},\boldsymbol{\rho}_{h}\right],\label{eq:b_eta}
\end{equation}
\begin{equation}
\boldsymbol{b}_{\nu}(t_{h},\boldsymbol{\rho}_{h})=\frac{i}{2\hbar}\left\{ \sigma_{z},\boldsymbol{\rho}_{h}\right\} ,\label{eq:b_nu}
\end{equation}
and $H_{h}=H\left(t_{h}\right)$. The noises are expressed as weighted
sums of white noise random numbers using the discretised form of Eqs.
\eqref{eq:noise-eta-eta}-\eqref{eq:noise-eta-mu}, where each white
noise $x_{i}$ and $\overline{x}_{j}$ is expressed as $x_{j}(t)dt\rightarrow dW^{j}(t)$
and $\overline{x}_{j}(\tau)d\tau\rightarrow d\overline{W}{}^{j}(\tau)$,
the overbar once more denoting a function of imaginary time $\tau$
and indices $j=$1,2,3 referring to specific white noises to enforce
the necessary correlations. For example, the $\eta_{\eta}$ noise
is expressed as 
\begin{equation}
\eta_{\eta}(t_{h})=\sum_{n=-N}^{N}G_{\eta\eta}(t_{n})dW^{1}(t_{h}-t_{n})
\end{equation}
with $j=1$. Here we shall use the index $n$ ranging between $-N$
and $N$ to denote discretised real time integrations with $t_{n}=n\,dt$,
and the index $m$ between $-M$ and $M$ for the integration in imaginary
time, $\tau_{m}=m\,d\tau$. The inverse Fourier transforms of the
filtering kernels, Eqs. \eqref{eq:filter-eta-eta}-\eqref{eq:filter-mu-mu},
and any other numerical prefactors, can freely be absorbed into the
diffusion function, giving them an additional index $n$ or $m$ associated
with the appropriate sum over time. This transforms the right hand
side of Eq. \eqref{eq:ESLE_ab} into the compact form $\sum_{j}B^{kj}(t_{h},\boldsymbol{\rho}_{h})dW_{t}^{j}$.
The rows $\mathbf{B}^{k}$ of the matrix $\boldsymbol{B}(t_{h},\boldsymbol{\rho}_{h})=\left\{ B_{h}^{kj}\right\} $
form vectors, each associated with one of the elements $\rho^{k}$
of the density vector, and are composed of the following components:
\[
\mathbf{B}^{k}(t_{h},\boldsymbol{\rho}_{h})=\left[\bigg(\underbrace{b_{\eta}^{k}G_{\eta\eta}(t_{n})dt}_{-N\leq n\leq N;\ j\in dW^{1}}\bigg)\quad\bigg(\underbrace{b_{\eta}^{k}G_{\eta\nu}(t_{n})dt+ib_{\nu}^{k}G_{\nu\eta}(t_{n})dt}_{-N\leq n\leq N;\ j\in dW^{2}}\bigg)\quad\bigg(\underbrace{ib_{\eta}^{k}G_{\eta\nu}(t_{n})dt+b_{\nu}^{k}G_{\nu\eta}(t_{n})dt}_{-N\leq n\leq N;\ j\in dW^{3}}\bigg)\right.
\]
\begin{equation}
\left.\quad\bigg(\underbrace{b_{\eta}^{k}G_{\eta\mu}(t_{h},\tau_{m})dt}_{-M\leq m\leq M;\ j\in d\overline{W}{}^{2}}\bigg)\quad\bigg(\underbrace{ib_{\eta}^{k}G_{\eta\mu}(t_{h},\tau_{m})dt}_{-M\leq m\leq M;\ j\in d\overline{W}{}^{3}}\bigg)\right]\label{eq:B-vector}
\end{equation}
$\mathbf{B}^{k}$ contains five sets of elements, each associated
with a different white noise. The index $j$ identifies the Wiener
increment of the appropriate white noise, and within each set of elements
the indices $n$ and $m$ run across real and imaginary times, respectively.

For each $t_{h}$, the increments $dW^{j}$ form a $2(2M+1)+3(2N+1)$
long vector $d\mathbf{W}$, elements of which are ordered in the same
way as inside the vector $\mathbf{B}^{k}$ above: 
\begin{align}
d\mathbf{W}_{h} & =\Bigg[\bigg(\underbrace{dW^{1}(t_{h}-t_{n})}_{-N\leq n\leq N}\bigg)\quad\bigg(\underbrace{dW^{2}(t_{h}-t_{n})}_{-N\leq n\leq N}\bigg)\quad\bigg(\underbrace{dW^{3}(t_{h}-t_{n})}_{-N\leq n\leq N}\bigg)\quad\bigg(\underbrace{d\overline{W}{}^{2}(\tau_{m})}_{-M\leq m\leq M}\bigg)\quad\bigg(\underbrace{d\overline{W}{}^{3}(\tau_{m})}_{-M\leq m\leq M}\bigg)\Bigg].\label{eq:W-vector}
\end{align}
These notations enable us to refer to either of the five sets of terms
in the sum $\sum_{j}B^{kj}(t_{h},\boldsymbol{\rho}_{h})dW_{h}^{j}$
by the particular family of the noise increments, e.g. $dW^{1}$ or
$d\overline{W}{}^{2}$, as is also indicated underneath each term
in Eq. \eqref{eq:B-vector}. The ESLN now has the standard form of
Eq. \eqref{eq:coupled_manynoise_index}, and it is clear that there
are many white noises appearing in this Langevin equation. This justifies
the choice of using Stratonovich calculus since Itô calculus would
be punitively expensive.

To transition into the Stratonovich-Heun scheme, we have to calculate
the Stratonovich correction
\[
-\frac{1}{2}\sum_{lj}B^{lj}(t_{h},\boldsymbol{\rho}_{h})\frac{\partial B^{kj}(t_{h},\boldsymbol{\rho}_{h})}{\partial\rho_{h}^{l}}
\]
needed for the modified drift, Eq. \eqref{eq:modified_drift}, for
each family of the increments $j\in dW^{1}$, $dW^{2}$, $dW^{3}$
, $d\overline{W}{}^{2}$ and $d\overline{W}{}^{3}$. For $j\in dW^{1}$,
we have $B^{kj}=b_{\eta}^{k}G_{\eta\eta}(t_{m})dt$, and only 
\begin{equation}
\boldsymbol{b}_{\eta}=\frac{2i}{\hbar}\begin{pmatrix}0 & \rho_{h}^{12}\\
-\rho_{h}^{21} & 0
\end{pmatrix}\:\rightarrow\:\frac{2i}{\hbar}\left(\begin{array}{cccc}
0 & \rho_{h}^{2} & -\rho_{h}^{3} & 0\end{array}\right)^{T},\label{eq:bk}
\end{equation}
(where we have used both the $2\times2$ matrix and the 4-fold vector
notations) depends on the elements $\rho_{h}^{l}$ of $\boldsymbol{\rho}_{h}$,
so the derivatives $\partial b_{\eta}^{k}/\partial\rho_{h}^{l}$ are
easily calculated forming a $4\times4$ matrix 
\begin{equation}
\left(\frac{\partial b_{\eta}^{k}}{\partial\rho_{h}^{l}}\right)=\frac{2i}{\hbar}\left(\begin{array}{cccc}
0 & 0 & 0 & 0\\
0 & 1 & 0 & 0\\
0 & 0 & -1 & 0\\
0 & 0 & 0 & 0
\end{array}\right)\label{eq:dbdrho}
\end{equation}
with respect to indices $k,l$. Substituting these into the Stratonovich
correction, we obtain the following contribution from the $dW^{1}$
terms:
\begin{equation}
-\frac{1}{2}\sum_{l,j\in dW^{1}}B^{lj}(t_{h},\boldsymbol{\rho}_{h})\frac{\partial B^{kj}(t_{h},\boldsymbol{\rho}_{h})}{\partial\rho_{h}^{l}}=2\left(\frac{dt}{\hbar}\right)^{2}\begin{pmatrix}0 & \rho_{h}^{12}\\
\rho_{h}^{21} & 0
\end{pmatrix}\sum_{m=-M}^{M}G_{\eta\eta}^{2}(t_{m})\label{eq:Stratonovich_correction_1}
\end{equation}
where we have returned back to the matrix notations for clarity.

Similarly, for $j\in dW^{2}$, we have $B^{kj}=ib_{\nu}^{k}G_{\nu\eta}(t_{m})dt+b_{\eta}^{k}\delta_{m0},$
and only $\boldsymbol{b}_{\nu}$ and its derivative are left to calculate:
\begin{equation}
\boldsymbol{b}_{\nu}=\frac{i}{\hbar}\begin{pmatrix}\rho_{h}^{11} & 0\\
0 & -\rho_{h}^{22}
\end{pmatrix}\:\rightarrow\:\left(\begin{array}{cccc}
\rho_{h}^{1} & 0 & 0 & -\rho_{h}^{4}\end{array}\right),
\end{equation}
\begin{equation}
\left(\frac{\partial b^{k}{}_{\nu}}{\partial\rho_{h}^{l}}\right)=\frac{i}{\hbar}\left(\begin{array}{cccc}
1 & 0 & 0 & 0\\
0 & 0 & 0 & 0\\
0 & 0 & 0 & 0\\
0 & 0 & 0 & -1
\end{array}\right)
\end{equation}
and the appropriate contribution to the correction from $dW^{2}$
noises is 
\begin{equation}
-\frac{1}{2}\sum_{l,j\in dW^{2}}B^{lj}(t_{h},\boldsymbol{\rho}_{h})\frac{\partial B^{kj}(t_{h},\boldsymbol{\rho}_{h})}{\partial\rho_{h}^{l}}=-\frac{1}{2}\left(\frac{dt}{\hbar}\right)^{2}\begin{pmatrix}\rho_{h}^{11} & 0\\
0 & \rho_{h}^{22}
\end{pmatrix}\sum_{m=-M}^{M}G_{\nu\eta}^{2}(t_{m})+\frac{2}{\hbar^{2}}\begin{pmatrix}0 & \rho_{h}^{12}\\
\rho_{h}^{21} & 0
\end{pmatrix}\label{eq:Statanovich_correction-2}
\end{equation}
In the same way, the correction for $j\in dW^{3}$ is found to be
identical to the correction\eqref{eq:Statanovich_correction-2} for
$j\in dW^{2}$ but with the opposite sign such that they exactly cancel.

For the Wiener increments associated with white noises in imaginary
time, $j\in d\overline{W}{}^{2}$ and $j\in d\overline{W}{}^{3}$,
inspection of the elements of the $\boldsymbol{B}$ matrix, Eq. \eqref{eq:B-vector},
reveals that the terms associated with $d\overline{W}{}^{3}$ are
just $i$ times the terms associated with $d\overline{W}{}^{2}$.
The Stratonovich correction for these terms will thus be identical
apart from a minus sign coming from $i^{2}=-1$ in $d\overline{W}{}^{3}$,
and they will also exactly cancel. Thus only terms from $j\in dW^{1}$
contribute to the modified drift in Eq. \eqref{eq:modified_drift},
\begin{equation}
\tilde{\boldsymbol{a}}(t_{h},\boldsymbol{\rho}_{h})=\boldsymbol{a}(t_{h},\boldsymbol{\rho}_{h})+2\left(\frac{dt}{\hbar}\right)^{2}\begin{pmatrix}0 & \rho_{h}^{12}\\
\rho_{h}^{21} & 0
\end{pmatrix}\sum_{m=-M}^{M}G_{\eta\eta}^{2}(t_{m}).\label{eq:real_time_modified_drift}
\end{equation}
It is then straightforward to convert this into the corresponding
corrections for the spin $S_{x},S_{y},S_{z}$ and the trace $S_{\text{Tr}{\boldsymbol{\rho}}}$,
Eqs. \eqref{eq:xspin}-\eqref{eq:trace_evolution1}, yielding, respectively,
\begin{equation}
S_{x}(t_{h})=2\left(\frac{dt}{\hbar}\right)^{2}\sum_{m=-M}^{M}G_{\eta\eta}^{2}(t_{m})\sigma_{x}(t_{h})\label{eq:Sx}
\end{equation}
\begin{equation}
S_{y}(t_{h})=2\left(\frac{dt}{\hbar}\right)^{2}\sum_{m=-M}^{M}G_{\eta\eta}^{2}(t_{m})\sigma_{y}(t_{h})\label{eq:Sy}
\end{equation}
\begin{equation}
S_{z}(t_{h})=S_{\text{Tr}{\boldsymbol{\rho}}}(t_{h})=0,\quad\forall h.\label{eq:SzTr}
\end{equation}

\subsection{Imaginary Time Propagation}

It is straightforward to repeat the same procedure for thermalisation,
Eq. \eqref{eq:ESLE_thermalisation}, 
\begin{equation}
\overline{\rho}_{h+1}^{k}=\overline{\rho}_{h}^{k}-H_{0}\overline{\rho}_{h}^{k}d\tau+\sigma_{z}\overline{\rho}_{h}^{k}\mu(\tau_{h})d\tau,
\end{equation}
where $H_{0}$ is the initial Hamiltonian at the beginning of the
real time evolution, and $h$ is now an index associated with the
imaginary time $\tau_{h}=h\,d\tau$. There is no additional complexity
here as compared to the real time evolution, so for expedience the
result for the modified drift is simply stated: 
\begin{equation}
\tilde{\boldsymbol{a}}(\tau_{h},\overline{\boldsymbol{\rho}}_{h})=-\left[H_{0}+\frac{1}{2}\left(\frac{d\tau}{\hbar}\right)^{2}\sum_{n=-N}^{N}G_{\mu\mu}^{2}(\tau_{n})\right]\overline{\boldsymbol{\rho}}_{h}.\label{eq:thermalisation_modified_drift}
\end{equation}

\end{document}